\documentclass[twoside,twocolumn,9pt]{article}
\usepackage{extsizes}
\usepackage[super,sort&compress,comma]{natbib} 
\usepackage[version=3]{mhchem}
\usepackage[left=1.5cm, right=1.5cm, top=1.785cm, bottom=2.0cm]{geometry}
\usepackage{balance}
\usepackage{widetext}
\usepackage{times,mathptmx}
\usepackage{sectsty}
\usepackage{graphicx} 
\usepackage{lastpage}
\usepackage[format=plain,justification=raggedright,singlelinecheck=false,font={stretch=1.125,small,sf},labelfont=bf,labelsep=space]{caption}
\usepackage{float}
\usepackage{fancyhdr}
\usepackage{fnpos}
\usepackage[english]{babel}
\usepackage{array}
\usepackage{droidsans}
\usepackage{charter}
\usepackage[T1]{fontenc}
\usepackage[utf8]{inputenc}
\usepackage[usenames,dvipsnames]{xcolor}
\usepackage{setspace}
\usepackage[compact]{titlesec}
\usepackage{blindtext}
\newcommand{\overbar}[1]{\mkern 1.5mu\overline{\mkern-1.5mu#1\mkern-1.5mu}\mkern 1.5mu}
\usepackage{adjustbox}
\usepackage{url}
\usepackage{listings}
\usepackage{hyperref}

\usepackage{epstopdf}

\definecolor{cream}{RGB}{222,217,201}

\begin{document}

\pagestyle{fancy}
\thispagestyle{plain}
\fancypagestyle{plain}{

\fancyhead[C]{\includegraphics[width=18.5cm]{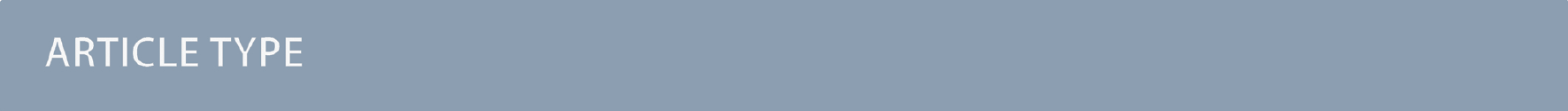}}
\fancyhead[L]{\hspace{0cm}\vspace{1.5cm}\includegraphics[height=30pt]{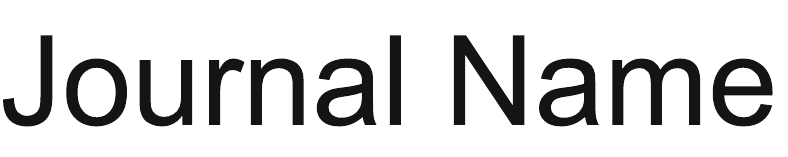}}
\fancyhead[R]{\hspace{0cm}\vspace{1.7cm}\includegraphics[height=55pt]{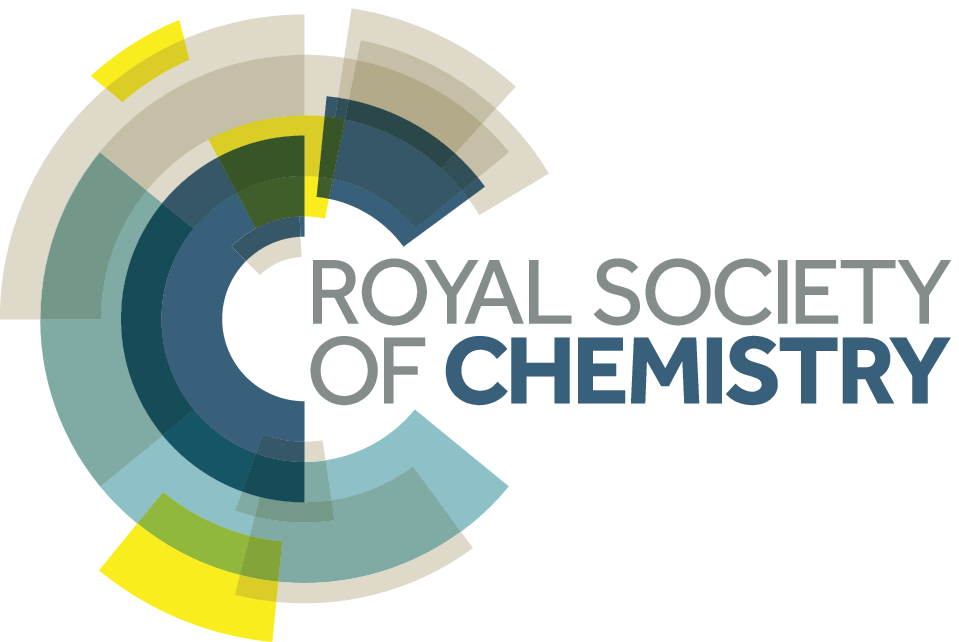}}
\renewcommand{\headrulewidth}{0pt}
}

\makeFNbottom
\makeatletter
\renewcommand\LARGE{\@setfontsize\LARGE{15pt}{17}}
\renewcommand\Large{\@setfontsize\Large{12pt}{14}}
\renewcommand\large{\@setfontsize\large{10pt}{12}}
\renewcommand\footnotesize{\@setfontsize\footnotesize{7pt}{10}}
\makeatother

\renewcommand{\thefootnote}{\fnsymbol{footnote}}
\renewcommand\footnoterule{\vspace*{1pt}%
\color{cream}\hrule width 3.5in height 0.4pt \color{black}\vspace*{5pt}} 
\setcounter{secnumdepth}{5}

\makeatletter 
\renewcommand\@biblabel[1]{#1}            
\renewcommand\@makefntext[1]%
{\noindent\makebox[0pt][r]{\@thefnmark\,}#1}
\makeatother 
\renewcommand{\figurename}{\small{Fig.}~}
\sectionfont{\sffamily\Large}
\subsectionfont{\normalsize}
\subsubsectionfont{\bf}
\setstretch{1.125} 
\setlength{\skip\footins}{0.8cm}
\setlength{\footnotesep}{0.25cm}
\setlength{\jot}{10pt}
\titlespacing*{\section}{0pt}{4pt}{4pt}
\titlespacing*{\subsection}{0pt}{15pt}{1pt}

\fancyfoot{}
\fancyfoot[LO,RE]{\vspace{-7.1pt}\includegraphics[height=9pt]{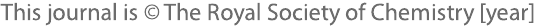}}
\fancyfoot[CO]{\vspace{-7.1pt}\hspace{13.2cm}\includegraphics{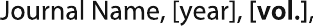}}
\fancyfoot[CE]{\vspace{-7.2pt}\hspace{-14.2cm}\includegraphics{head_foot/RF}}
\fancyfoot[RO]{\footnotesize{\sffamily{1--\pageref{LastPage} ~\textbar  \hspace{2pt}\thepage}}}
\fancyfoot[LE]{\footnotesize{\sffamily{\thepage~\textbar\hspace{3.45cm} 1--\pageref{LastPage}}}}
\fancyhead{}
\renewcommand{\headrulewidth}{0pt} 
\renewcommand{\footrulewidth}{0pt}
\setlength{\arrayrulewidth}{1pt}
\setlength{\columnsep}{6.5mm}
\setlength\bibsep{1pt}

\makeatletter 
\newlength{\figrulesep} 
\setlength{\figrulesep}{0.5\textfloatsep} 

\newcommand{\topfigrule}{\vspace*{-1pt}%
\noindent{\color{cream}\rule[-\figrulesep]{\columnwidth}{1.5pt}} }

\newcommand{\botfigrule}{\vspace*{-2pt}%
\noindent{\color{cream}\rule[\figrulesep]{\columnwidth}{1.5pt}} }

\newcommand{\dblfigrule}{\vspace*{-1pt}%
\noindent{\color{cream}\rule[-\figrulesep]{\textwidth}{1.5pt}} }

\makeatother

\newcommand*{\citen}[1]{%
  \begingroup
    \romannumeral-`\x 
    \setcitestyle{numbers}%
    \cite{#1}%
  \endgroup
}

\twocolumn[
  \begin{@twocolumnfalse}
\vspace{3cm}
\sffamily
\begin{tabular}{m{4.5cm} p{13.5cm} }

\includegraphics{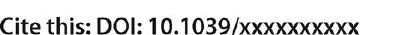} & \noindent\LARGE{\textbf{In-Silico Characterization of Nanoparticle Catalysts$^\dag$}} \\
\vspace{0.3cm} & \vspace{0.3cm} \\

 & \noindent\large{Björn Kirchhoff,$^{\ast}$\textit{$^{, a}$} Christoph Jung,\textit{$^{a,b,c}$} Daniel Gaissmaier,$^{a,b,c}$ Laura Braunwarth,$^{a}$ Donato Fantauzzi and Timo Jacob$^{\ast}$\textit{$^{, a,b,c}$}} \\

\includegraphics{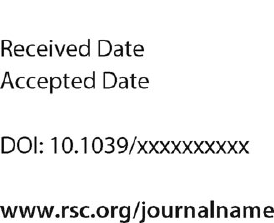} & \noindent\normalsize{Nanoparticles (NPs) make for intriguing heterogeneous catalysts due to their large active surface area and excellent and often size-dependent catalytic properties that emerge from a multitude of chemically different surface reaction sites. NP catalysts are, in principle, also highly tunable: even small changes to the NP size or surface facet composition, doping with heteroatoms, or changes of the supporting material can significantly alter their physicochemical properties. Because synthesis of size- and shape-controlled NP catalysts is challenging, the ability to computationally predict the most favorable NP structures for a catalytic reaction of interest is an in-demand skill that can help accelerate and streamline the material optimization process. Fundamentally, simulations of NP model systems present unique challenges to computational scientists. Not only must considerable methodological hurdles be overcome in performing calculations with hundreds to thousands of atoms while retaining appropriate accuracy to be able to probe the desired properties. Also, the data generated by simulations of NPs are typically more complex than data from simulations of, for example, single crystal surface models, and therefore often requires different data analysis strategies. To this end, the present work aims to review analytical methods and data analysis strategies that have proven useful in extracting thermodynamic trends from NP simulations.}\\

\end{tabular}

 \end{@twocolumnfalse} \vspace{0.6cm}

  ]

\renewcommand*\rmdefault{bch}\normalfont\upshape
\rmfamily
\section*{}
\vspace{-1cm}


\footnotetext{\textit{$^{a}$~Institute of Electrochemistry, Ulm University, Albert-Einstein-Allee 47, 89081 Ulm, Germany. Fax: +49 731 5025409; Tel: +49 731 5025401; E-mail: nawi.ec@uni-ulm.de}}
\footnotetext{\textit{$^{b}$~Helmholtz-Institute Ulm (HIU) Electrochemical Energy Storage, Helmholtz-Straße 16, 89081 Ulm, Germany.}}
\footnotetext{\textit{$^{c}$~Karlsruhe Institute of Technology (KIT), P.O. Box 3640, 76021 Karlsruhe, Germany.}}

\footnotetext{\dag~Electronic Supplementary Information (ESI) available: An online tutorial for implementing many of the presented methods is available via \href{https://bjk24.gitlab.io/in-silico-review/intro.html}{https://bjk24.gitlab.io/in-silico-review/intro.html}.}




\section{Introduction}

Deployment of heterogeneous catalysts in nanoparticulate form has various benefits. Not only do nanoparticles (NPs) promise high mass activity due to the more favorable surface-to-volume ratio compared to catalysts with grain sizes in the micro- or millimeter range. NPs can also express heightened or outright different catalytic properties compared to the bulk material as a result of finite- and quantum-size effects and due to the availability of a plethora of chemically distinct surface reaction sites. However, synthesis of shape- and size-controlled NPs is challenging, and Ostwald ripening as well as other degradation effects can impact the longevity of NP catalysts.\cite{meier2014} In order to streamline material optimization cycles, interest has therefore been growing in computational approaches that can predict favorable catalyst candidate structures for a reaction of interest. More and more, computational science is asked to establish structure-activity relationships for model NP catalysts in the 1 to 5~nm range and to investigate degradation mechanisms.

Historically, computational investigation of NP properties was usually carried out by performing density functional theory (DFT) calculations of single crystal model surfaces that correspond to the surface facets of a NP of interest.\cite{tian2008,kaghazchi_bridging_2009,tritsaris_atomic-scale_2011,zhu_roughening_2013,ouyang2016} However, results from surface models are often not transferable to NPs for several reasons. Firstly, the number of chemically different surface sites on a NP is larger than on single crystal surface models. In particular, the highly reactive, undercoordinated edge and vertex sites are hard to represent using a surface model.\cite{yudanov_metal_2002} Furthermore, when using stepped model surfaces to mimic the lower coordination of NP edge sites, one finds that such models contain both convex and concave surface structures while NPs typically only contain one type.\cite{callevallejo2017} Another disparity between surface models and actual NPs is that quantum-size effects can affect the properties of clusters and small NPs in non-systematic ways.\cite{volokitin_quantum-size_1996,kleis2011} In fact, not all properties observed for NPs will converge to the bulk limit as the system size increases.\cite{vines_understanding_2014} Surface models are therefore unsuitable to study certain NP properties, even if the surface model is used as a stand-in for very large NPs. Finally, investigation of catalyst degradation processes using surface models and DFT can be arduous given the computational limitations with regards to system size and time scale that such processes typically occur at.

Nevertheless, the surface model approach remained the default option for a long time because accurate simulations of NPs with hundreds or thousands of atoms are computationally demanding. Recent advances in computational methodology have changed this circumstance. Force field-based methods such as ReaxFF\cite{van_duin_reaxff_2001,han_development_2016} or semi-empirical methods such as xTB\cite{grimme_robust_2017} and Sutton-Chen type potentials\cite{sutton_long-range_1990,qi_melting_2001,tian_platinum_2008,huang2011} have brought down costs for calculations of NPs significantly while retaining reasonable accuracy. Similarly, highly scalable codes enable electronic structure calculations of model sizes previously thought inaccessible using DFT. Advances in the field of hybrid simulations such as the QMMM ansatz also push the limits of system size.\cite{warshel1976,thole1980,field1990,lu_qmmm_2016,dohn_multiscale} 

With computational costs for NP simulations brought down, one important question remains: how best to extract useful information from the model systems and simulations? Analyzing properties and the reactivity of NPs is not as straight forward as analyzing a single crystal surface for example. The abundance of different binding sites will naturally lead to noisier results. Oftentimes, molecular dynamics (MD) or Monte Carlo (MC) based simulations are used to study NP model systems. Such simulations can create large data sets that require specialized analysis to extract the desired information.

This work aims to review methods of data analysis that we found useful in our research in this field\cite{kirchhoff2019,braunwarth_exploring_2020,Kirchhoff2022} to extract information about thermodynamic properties and structure-activity relationships from simulations of NP catalysts. Exemplary studies applying the presented methods are highlighted when appropriate, without the intention to be exhaustive in this regard.

Many of the methods discussed in this review will be illustrated using atomic configurations of 3~nm octahedral Pt NP structures in various states of oxidation. These structures are taken from an openly accessible data set available under the DOI \href{https://doi.org/10.5281/zenodo.6322004}{10.5281/zenodo.6322004} which our group recently published alongside a study comparing electrochemical oxidation trends of Pt NPs of various shapes.\cite{Kirchhoff2022} Additionally, we have made available a git repository of a Jupyter book that serves as a tutorial on how to implement many of the methods presented below using Python3: \href{https://bjk24.gitlab.io/in-silico-review/intro.html}{https://bjk24.gitlab.io/in-silico-review/intro.html}. The code examples are written to directly interact with the data set for oxidized octahedral NPs. More information on the required setup to follow the tutorial and execute the code for yourself can be found on the website linked above.

Of course, almost all of the presented methods are already very efficiently implemented in existing software packages for daily use. However, such programs are typically closed-source or at least hard to decipher due to radical optimization for speed. We therefore encourage readers, who we anticipate are scientists first and programmers second, to run and modify our more straight forward test script suite to familiarize themselves with the implementation of these methods and with code-driven data analysis in general.

The remainder of this review is structured as follows: section 2 shortly summarizes open-source software solutions that enable users to generate atomic configurations of NPs. Section 3 is concerned with analytical methods aimed at extracting information from individual structures while section 4 presents methods aimed at analyzing entire data sets. Some concluding remarks are given in section 5. All figures original to this work used the Visual Molecular Dynamics (VMD) software\cite{HUMP96} and the Tachyon renderer\cite{STON1998} to visualize atomic configurations. If not indicated otherwise, Pt atoms are grey while oxygen atoms are red. 

\section{Generating Nanoparticle Model Systems}

Many commercial computational chemistry software suites allow users to generate NP configurations. Here, three open-source alternatives are mentioned to reduce the activation barrier for newcomers.
\begin{itemize}
    \item Atomic Simulation Environment (ASE): ASE is a python3-based framework for molecular and materials simulations tgat interfaces with many commercial and open-source computational codes.\cite{larsen2017} The \verb|ase.cluster| class provides various means for generating NPs based on user-provided information about the lattice type and constants as well as Miller indices of the surfaces that should be cut. Pre-configured methods exist to generate typical NP shapes such as octahedra. A method for Wulff constructions exists as well which requires users to supply the formation energy of different types of surface facets. The method will then create a NP shape that minimizes the overall surface energy.
    \item WulffPack: WulffPack is a powerful python3 package used to generate NPs using Wulff construction.\cite{Rahm2020} WulffPack features built in methods to visualize the generated NPs and interfaces with ASE to export atomic coordinates.
    \item Nanocut: nanocut is a python3 package that can be used to generate NPs by providing cell and Miller index information. The program can also generate periodic surfaces and crystals, nanotubes, and nanowires.\cite{aradi_2019} The ability to chain sequences of cutting procedures allows users to realize unusual shapes. At the time of writing, development appears to have halted. Nanocut may still be worth a look for users who find other alternatives too restrictive as compatibilty issues should be quick to fix due to the open-source nature of the project.
\end{itemize}

\section{Analyzing Single Structures}

\subsection{Evaluating Surface Coverage of Nanoparticles} \label{s:coverage}

In computational surface science, coverage is typically given in terms of monolayers, \textit{i.e.} with respect to the number of atoms within the top layer of the underlying periodic surface model. This is often a convenient measure because it directly relates to single-crystal experiments for example. In case of NPs, however, a monolayer is somewhat ill-defined due to the inhomogeneous nature of the available surface sites. Experimental data on NPs, for example from X-ray spectroscopy methods, can give insight into the relative occurrence of different chemical species or oxidation states in a material. Hence, giving coverage as a ratio of $x = \text{number of adatoms} / \text{number of substrate atoms}$ is found to be more useful.\cite{senftle2013,senftle2014-1,senftle2014,kirchhoff2019} Senftle and co-workers for example have characterized the degree of oxidation of a Pd NP via its O:Pd ratio.\cite{senftle2013} This method of designating coverage will be used throughout this review.

\subsection{Coordination Number-Based Approaches} \label{sec:cns}

\subsubsection{Classical Coordination Numbers}

The classical coordination number ($CN$) of a given atom is calculated by counting its nearest-neighbor atoms. $CN$s have been used to classify adsorption positions on model surfaces for decades.\cite{robertson1991} The usage of $CN$s to address adsorption sites on NPs is advantageous compared to the commonly used way of referring to adsorption sites on extended surfaces by means of the underlying surface composition (\textit{fcc}, \textit{hcp}, \textit{threefold}, \textit{etc.}) simply due to the sheer number of different available binding sites.

Moreover, $CN$s have been shown to be useful as a descriptor for catalytic activity. In 2009, Jiang and co-workers showed that the $CN$s of surface atoms on small NPs linearly correlate with metal \textit{d}-band centers,\cite{hammer1995} priming $CN$s as descriptors for reactivity.\cite{jiang2009} Mpourmpakis and co-workers exploited this relationship further: they used $CN$s to characterize the interactions between small gold clusters (\ce{Au16} to \ce{Au45}) and CO molecules.\cite{mpourmpakis2010} From these results, the group extrapolated a transferable concept that uses the $CN$ and the bonding angle of an adspecies on the particle surface as descriptors for the binding energy. This approach therefore allows for the prediction of NP reactivity without requiring expensive electronic structure calculations. In a 2015 follow up, the group introduced a refined descriptor approach based on the relationship between CO binding energy and surface $CN$ that takes into account interdependence of NP structure and adsorbate coverage.\cite{taylor2015}

Two interesting applications are highlighted in the following to further illustrate the usefulness of $CN$s for catalysis research. Using O and CO molecules as probes, Kleis \textit{et al.} employed $CN$s to identify the point of transition at which the adsorption properties on cuboctahedral Au NPs will converge to extended surface behavior.\cite{kleis2011} By relating adsorption energy values to $CN$s, the group was able to show that for particles with more than 561 atoms (diameter of \textit{ca.} 2.7 nm), the adsorption energy on a (111)-indexed Au NP facet becomes invariant against the extent of the facet and is close to the number obtained for an extended Au(111) model surface. In another example, Ouyang \textit{et al.} chose a Au(997) surface to simulate a NP system and used $CN$s to study O and NO adsorption.\cite{ouyang2016} The group identified the relative activity of different adsorbed oxygen species towards NO reduction which they found to be "island O" $>$ "terrace O" $>$ "O near step-edge sites".

\begin{figure}[htbp]
\centering
    \begin{minipage}{0.45 \linewidth}
        \centering
        \includegraphics[width=\linewidth]{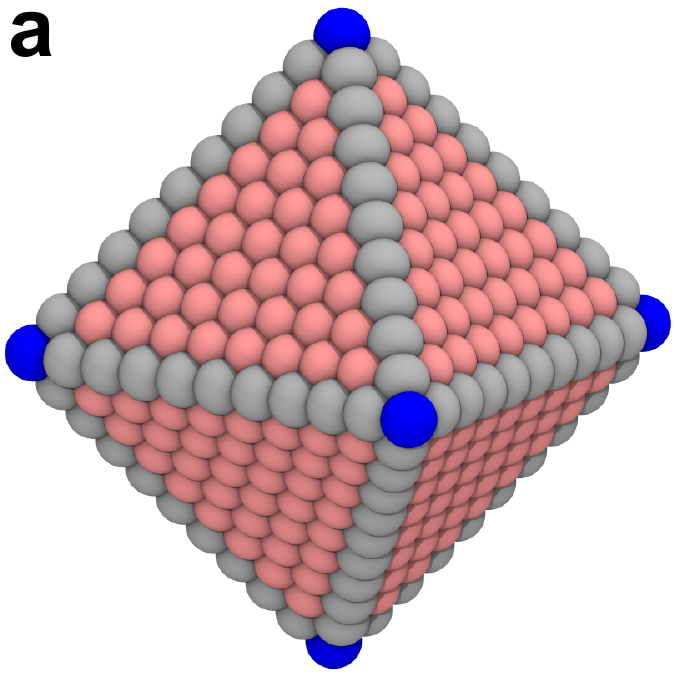}
    \end{minipage}\hspace{0.3cm}
    \begin{minipage}{0.45 \linewidth}
        \centering
        \includegraphics[width=\linewidth]{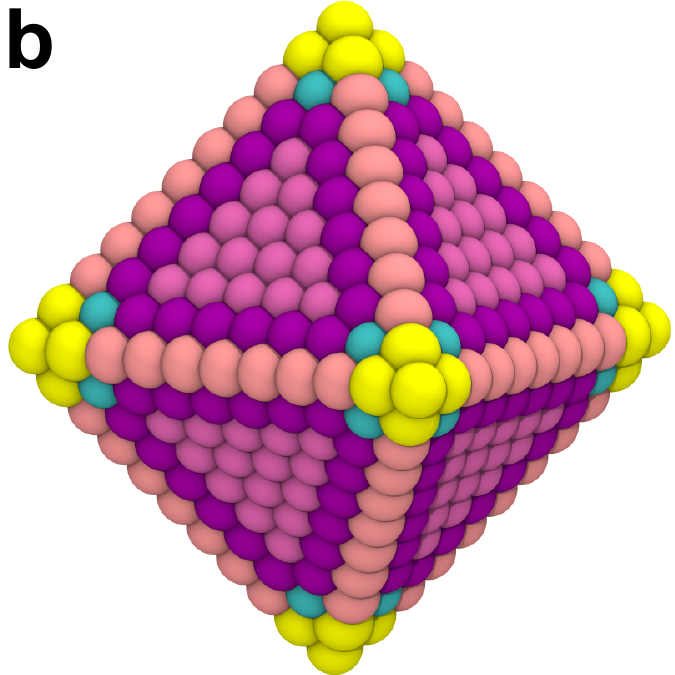}
    \end{minipage}\\[0.3cm]
    \begin{minipage}{0.45 \linewidth}
        \centering
        \includegraphics[width=\linewidth]{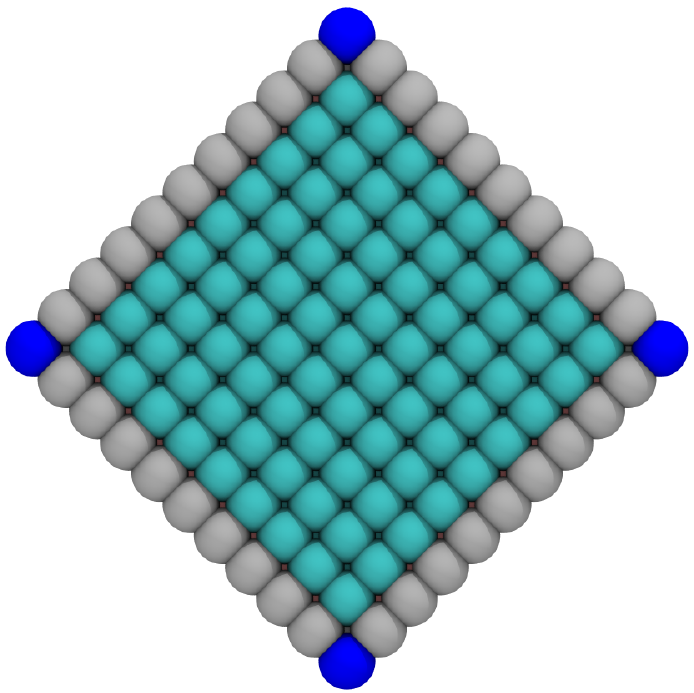}
    \end{minipage}\hspace{0.3cm}
    \begin{minipage}{0.45 \linewidth}
        \centering
        \includegraphics[width=\linewidth]{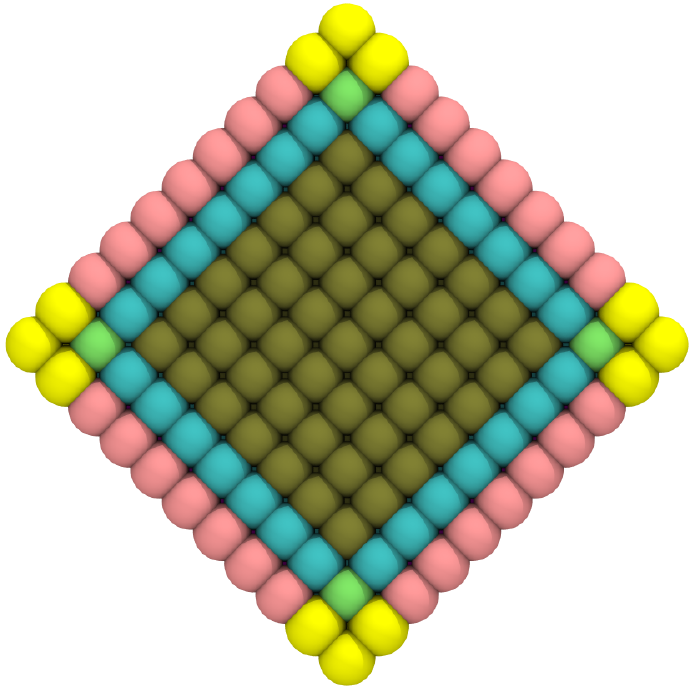}
    \end{minipage}
  \caption{Comparison of \textbf{a} classical coordination number analysis and \textbf{b} generalized coordination number analysis of a 3~nm octahedral Pt NP. Top: outside view. Bottom: cross section. Different colors indicate different coordination numbers.}
  \label{fgr:ccn_gcn}
\end{figure}

A noteworthy limitation of $CN$s is their low sensitivity for microscopic features. For example, the ABC-layered structure of a Pt(111) surface gives rise to two different threefold adsorption sites, HCP and FCC. It is well known that these two sites interact differently with certain adspecies such as oxygen; however, since $CN$s only take into account the first coordination sphere, HCP and FCC sites are assigned the same $CN$. Furthermore, when $CN$s are used with NPs, they do not capture edge or kink site effects. As shown in Fig. \ref{fgr:ccn_gcn}\textbf{A}, all facet atoms barring those located exactly on the edges and vertices are assigned values of the corresponding single crystal surface.
The study by Kleis \textit{et al.} discussed above established that a certain particle size is required before facet atoms express properties analogous to the corresponding single crystal surface.\cite{kleis2011} $CN$s as a categorization tool are too coarse-grained to capture such effects. This limitation is further exemplified in the cross section view in Fig. \ref{fgr:ccn_gcn}\textbf{A}. The cross section illustrates that $CN$s immediately converge to the bulk value of an \textit{fcc} crystal at the first layer below the surface. This step-like change does not reflect the more gradual --- albeit still fast --- decay of surface properties with increasing depth.

A code example for calculating CNs from an atomic configuration can be found in the supporting git repository in \verb|classicalCN.py|.

\subsubsection{Generalized Coordination Numbers}

Generalized coordination numbers ($\overbar{CN}$s) were introduced by Calle-Vallejo and co-workers in 2014.\cite{callevallejo2014} The $\overbar{CN}$ of a specific atom constitutes the average of the $CN$s of its nearest-neighbor atoms, see Equation \eqref{gcns}:
\begin{equation}
    \overbar{CN}(i) = \sum_{j=1}^{n_j} \frac{CN(j)}{CN_{\text{max}}}\, . \label{gcns}
\end{equation}
Here, $CN(j)$ are the classical coordination numbers of neighboring atoms $j$ and $CN_{\text{max}}$ is a weighing factor. $CN_{\text{max}}$ defines the largest possible coordination number (12 for an FCC crystal, 9 for a BCC crystal, \textit{etc}.). Contrary to CNs, $\overbar{CN}$s are distinct for different types of three-fold adsorption sites or bridge positions on ABC-layered surfaces. Similarly, $\overbar{CN}$s give a more nuanced picture of the different adsorption sites on NPs, see Fig. \ref{fgr:ccn_gcn}\textbf{B}.

Calle-Vallejo and co-workers have shown the usefulness of $\overbar{CN}$s in several studies. The group used $\overbar{CN}$s to study the interplay between adsorption energetics and adsorbate-induced core-shell deformation.\cite{callevallejo2014_2} They show that the deformation cost is site-independent and can be assumed constant for a specific adspecies. In 2015, the group used $\overbar{CN}$s to devise a relationship for adsorption properties of oxygen species on transition-metal surfaces that does not require expensive electronic structure calculations.\cite{callevallejo2015} The relationship they propose relies on knowledge of the valence of the adatom, which can be obtained from electron-counting rules, and the chemical environment of the active site, which they represent via $\overbar{CN}$s. The approach can be used to predict surface features that satisfy the energetic requirements of each intermediate of a given reaction mechanism, which Calle-Vallejo \textit{et al.} confirmed against benchmark DFT calculations. The group successfully applied this approach to inform the experimental modification of a Pt(111) surface which produced a material with threefold enhanced ORR activity.\cite{callevallejo2015_2} Tymoczko \textit{et al.} further used this approach to explain the heightened experimental HER activity of a Cu-Pt(111) near-surface alloy.\cite{tymoczko2016} 

Recently, Calle-Vallejo \textit{et al.} used $\overbar{CN}$s to clarify why simulation results from single crystal model surfaces are oftentimes not transferable to NPs.\cite{callevallejo2017} In particular, they note that Pt surfaces contain both convex and concave sites while typical NP systems like cuboctahedra contain only convex sites that are less active towards the ORR. They find that active sites with $\overbar{CN} > 7.5$ promote ORR activity, where $\overbar{CN} = 7.5$ corresponds to Pt atoms in an extended Pt(111) surface.

$\overbar{CN}$s are ill-defined in case of alloy catalysts for which not only the number but also the elemental nature of neighboring atoms are important to describe the properties of a specific site. Similarly, $\overbar{CN}$s cannot be used to study systems for which strain is the dominant factor governing chemical reactivity.

A code example for calculating $\overbar{CN}$s from an atomic configuration can be found in the git supporting repository as \verb|generalizedCN.py|.

\subsubsection{Orbitalwise Coordination Numbers}

In order to overcome the limitations of $\overbar{CN}$s outlined above, Ma and Xin extend the coordination number approach to orbitals. Orbitalwise coordination numbers ($CN^\alpha$, where $\alpha$ refers to the orbital $\alpha$ = $s$, $d$)\cite{ma2017} are based on the electronic structure by way of moment characteristics of the projected density of states. Ma and Xin define $CN^\alpha$ in Eq. \eqref{orbitalcn} as
\begin{equation}
    CN^{\alpha}_i = \frac{M^{\alpha}_{2,1}}{(t_{nn}^{\alpha,\infty})^2} = \frac{\sum\limits_j^{r_{ij} < r_c}(t_{ij}^{\alpha})^2}{(t_{nn}^{\alpha,\infty})^2}\, ,  \label{orbitalcn}
\end{equation}
where $M^{\alpha}_{2,1}$ is the second moment from the moments theorem\cite{gaspard1973} and $t_{nn}^{\alpha,\infty}$ is the sum over all two-center hopping integrals of an $\alpha$ = $s$ or $d$ electron to valence orbitals of nearest-neighbor atoms. $M^{\alpha}_{2,1}$ is sensitive to perturbations introduced by the local chemical environment of a reactive site, thus accounting for effects such as material strain. 

\begin{figure}[htbp]
    \centering
    \includegraphics[width=\linewidth]{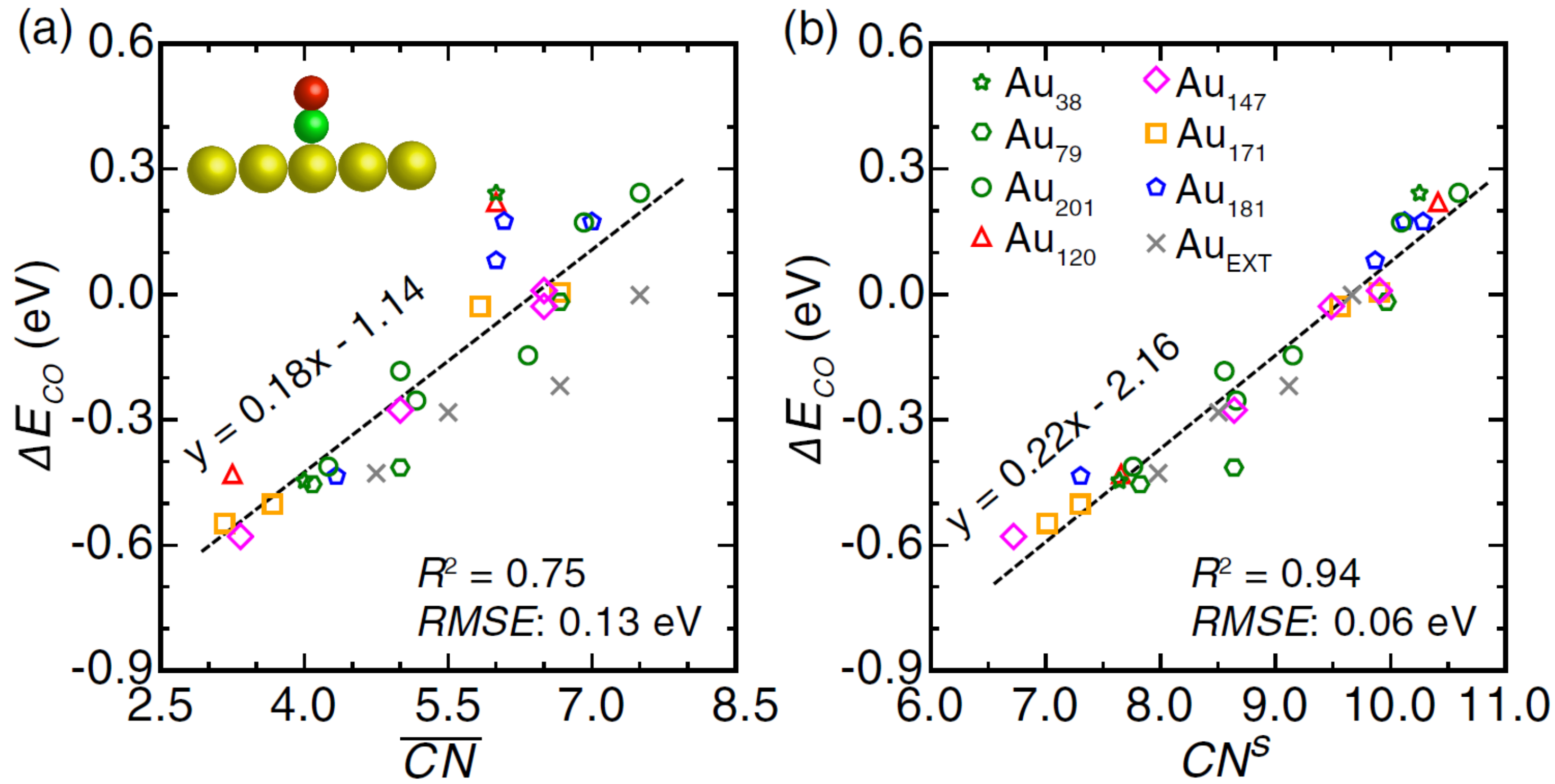}
    \caption{Comparison of CO adsorption energy trends on Au clusters and extended surface obtained with DFT and a (a) $\overbar{CN}$- or (b) $CN^\alpha$-based descriptor method. Reprinted with permission from Ma and Xin, \textit{Phys. Rev. Lett.}, 2017, \textbf{118}, 036101. Copyright 2017 by the American Physical Society.}
    \label{fgr:cnalpha}
\end{figure}

Ma and Xin tested the $CN^\alpha$ descriptor against $\overbar{CN}$s for CO adsorption on small Au clusters of 38 to 181 atoms, see Fig. \ref{fgr:cnalpha}.
The $CN^\alpha$-based model, which in this case relied on $s$ electron hopping integrals ($CN^s$) exclusively, is found to correlate better with DFT-PBE results ($R^2$ = 0.94) than the model based on $\overbar{CN}$s ($R^2$ = 0.75). Wang \textit{et al.}\cite{wang2018} used this approach to study larger Au NPs optimized using the embedded-atom method (EAM).\cite{daw_embedded-atom_1984} Adsorption energy results on individual sites were evaluated using $s$-orbital dependent coordination numbers. Wang \textit{et al.} further included a microkinetics model to predict CO oxidation rates as a function of particle size. They found improved agreement between experiments and the $CN^\alpha$ based model over using classical or $\overbar{CN}$s.

\subsubsection{Adjusted Coordination Numbers}

Even though $\overbar{CN}$s offer greater resolution than classic $CN$s, they can still be too coarse-grained for highly disordered systems such as transition metal oxides for which different oxidation states of surface metal atoms, surface defects, and other non-trivial electronic properties complicate the picture. Adjusted coordination numbers ($ACN$s) were proposed by Fung and co-workers in 2017 as a descriptor for transition metal oxides in particular.\cite{fung2017} Here, the $ACN$ of an oxygen atom located at the surface of the system is calculated based on its own coordination number, $CN_\text{O}$, and the sum over the coordination numbers of nearest-neighbor metal centers, $\sum CN_{\text{M}}$, as presented in Eq. \eqref{acns}:
\begin{equation}
    ACN = CN_\text{O} \cdot \lambda - \sum CN_\text{M}. \label{acns}
\end{equation}
The constant scaling factor $\lambda$ is related to the partial charge of the surface oxide. Therefore, conducting a partial charge analysis (see section \ref{s:pca}) of all relevant solid-state oxide crystal structures is necessary to calibrate the method for each new material. $ACN$s offer higher resolution than classic $CN$s or $\overbar{CN}$s and are therefore advantageous for describing the local chemical environment of low-coordination, surface-bound species such as oxides in more detail. In their publication, Fung \textit{et al.} calibrated $ACN$s for bulk \ce{Co3O4}, \ce{V2O3}, \ce{NiO}, and \ce{C-H} bond activation barriers.\cite{fung2017}

\subsection{Partial Charge Analysis} \label{s:pca}

\subsubsection{Mulliken}

Mulliken proposed the first scheme for partial charge analysis of molecules in 1955. \cite{mulliken1955_1,mulliken1955_2,mulliken1955_3,mulliken1955_4} The Mulliken scheme is based on the theory of representing molecular wavefunctions as linear combinations of atomic orbitals (LCAO) and is used to quantify the bonding character of the molecular orbital of a pair of two atoms. From LCAO theory, the molecular orbital of a pair of atoms can be described as\cite{zielinski2005}
\begin{equation}
    \psi_i = c_{ij}\, \phi_j + c_{ik}\, \phi_k, \label{eq:mulliken1}
\end{equation}
where $\psi_i$ is the molecular LCAO wavefunction, $\phi_{j,k}$ are the wavefunctions of atoms $j$ and $k$, and $c_{ij}$ and $c_{ik}$ are coefficients of the atomic orbitals. Squaring the wavefunction in Eq. \eqref{eq:mulliken1} gives the probability density. Normalization of the probability density gives the expression
\begin{equation}
    1 = c_{ij}^2 + c_{ik}^2 + c_{ij}\, c_{ik}\, S_{jk},
\end{equation}
where $S_{jk}$ is the overlap integral over the orbitals of atoms $j$ and $k$. Mulliken defined $c_{ij}^2$ and $c_{ik}^2$ as the atomic-orbital populations and $c_{ij}\, c_{ik}\, S_{jk}$ as the overlap population. The latter is < 0 for antibonding, 0 for nonbonding, and > 0 for bonding molecular orbitals. These quantities can be arranged in matrix form, the population matrix, where atomic-orbital populations constitute the diagonal and overlap populations the off-diagonal elements. Since applying this type of analysis to each pair of atoms in a system can lead to a large amount of data, different ways of post-treatment have been devised. The net population matrix, for example, is the sum of all population matrices for occupied orbitals; here, the diagonal elements give the total charge in each atomic orbital.

Mulliken charges suffer from a major limitation. Computational codes using the LCAO approach represent atoms through basis sets of different makeup and size. Basis sets may include varying numbers of basis functions to represent an orbital. The total atomic charge extracted via Mulliken's approach depends on the basis set size and has no basis set limit.\cite{reed1985} Mulliken occupancies typically do not sum up to the total number of electrons in the system and can even be negative in rare cases.\cite{reed1985} Finally, Mulliken analysis cannot be used with plane wave calculations directly.

\subsubsection{Hirshfeld} \label{s:hirshfeld}

While Mulliken charge analysis applies a basis-set based partitioning scheme, the Hirshfeld approach (1977) is based on separation of the electron density.\cite{hirshfeld1977} Here, the charge density is divided at each point between atom pairs, proportional to their respective contributions based on the free atoms. This strategy led to the byname of "stockholder" partition analysis. The partitioning procedure is based on earlier works by Politzer and Harris, refined by Hirshfeld to ensure that atomic fragments are always well defined.\cite{Politzer,hirshfeld1977} Comparison of the molecular electron density at a site $\overbar{r}$, $\rho^\text{mol}(\overbar{r})$, to the charge density of the "promolecule", $\rho^\text{pro}(\overbar{r})$, which is constructed from the individual non-interacting atomic densities, results in the deformation density $\rho_\text{d}(\overbar{r})$, as shown in Eq. \eqref{eq:hirshfeld1}:\cite{Saha2009}
\begin{equation}
    \rho_\text{d}(\overbar{r}) = \rho^\text{mol}(\overbar{r}) - \rho^\text{pro}(\overbar{r}) = \rho^\text{mol}(\overbar{r}) - \sum_{\alpha} p_{\alpha} (\overbar{r}-\overbar{R}_{\alpha}). \label{eq:hirshfeld1}
\end{equation}
$\sum\limits_{\alpha} p_{\alpha} (\overbar{r}-\overbar{R}_{\alpha})$ is the sum over the spherically averaged ground state electron densities of all free atoms $\alpha$, centered at the location of nucleus $\overbar{R}_{\alpha}$. For an uncharged molecule, the effective charge of an atom $q_{\alpha}$ can be obtained using Eq. \eqref{eq:hirshfeld2},
\begin{equation}
    q_{\alpha} = - \int \rho_\text{d}(\overbar{r}) \omega_{\alpha}(\overbar{r})d^3\overbar{r}. \label{eq:hirshfeld2}
\end{equation}
Here, the weighing function $ \omega_{\alpha}(\overbar{r})$ denotes the relative share that atom $\alpha$ contributes to the density at site $\overbar{r}$.

De Proft and co-workers extensively tested the Hirshfeld scheme on a series of functionalized organic molecules.\cite{proft2002} They note that while overall trends for charges are in good agreement with measurements, the magnitude of charges and dipoles is frequently underestimated. This conclusion is corroborated by other benchmark studies.\cite{choudhuri2020,manz2016,guerra2004} Furthermore, Hirshfeld charges can be affected unfavourably by atomic fragments exerting electron withdrawal or excess electrons in an atom's vicinity, as often observed by contradictory charges on hydrogen atoms.\cite{Saha2009,Davidson1992} Saha \textit{et al.} proposed refinements to overcome these limitations.\cite{Saha2009}

\subsubsection{Natural Population Analysis}

Natural population analysis (NPA) was introduced in the late 1980s by the Weinhold group\cite{reed1985,reed1985-1,reed1988} as a reaction to the strong basis set dependence of the Mulliken method. NPA uses natural atomic orbitals (NAOs)\cite{reed1983} which are based on natural orbitals (NOs) as introduced by L{\"o}wdin in the 1950s.\cite{lowdin1955,lowdin1955-1} NAOs do not depend on the basis set due to an occupation-dependent weighing factor which makes sure that unoccupied Rydberg states do not significantly impact the results of the method. Thus, unlike NOs, NAOs always remain close in shape to the original single-atom orbitals and electrons are localized close to the atomic centers for the purpose of counting occupations.

Unlike in case of the Mulliken approach, NPA produces only physically meaningful, non-negative occupations and the overall charge will sum up to the total number of electrons in the system. Weinhold and co-workers showed that NPA can significantly improve results over Mulliken and L{\"o}wdin analysis by varying the exponent of a \textit{2p} basis function in the SCF calculation of \ce{H2} and relating the resulting energy changes to charge results obtained from the different partial charge analysis methods.\cite{reed1988}

The NPA and Mulliken schemes share the limitation of not being directly applicable to plane wave calculations. Furthermore, dipole moments are overestimated significantly using NPA, which is discussed to be a result of the far-reaching tails of (diffuse) basis functions.\cite{guerra2004,martin2005}

\subsubsection{Bader (Atoms-In-Molecules)}

Bader charge analysis, also referred to as the atoms-in-molecules (AIM) method, was proposed in 1990\cite{bader1990,bader2004} and has been implemented in various codes over the years.\cite{popelier1994,stefanov1995,popelier1998,silvi2000,popelier2001,malcolm2003,malcolm2003-1,henkelman2006,sanville2007,tang2009} Bader analysis differs from the previously described methods in that it uses the electron density, thereby allowing both LCAO and plane waves methods to be used. Some implementations are, moreover, completely independent of the nuclear coordinates.\cite{henkelman2006,sanville2007,tang2009}

Fundamentally, Bader analysis separates the system by creating curved surfaces that run through minima in the charge density. The volume segments enclosed by these surfaces are referred to as Bader regions. The total charge of each Bader region is obtained by integration. Bader charge analysis has been shown to systematically overestimate charges and dipole moments.\cite{guerra2004,martin2005,choudhuri2020}

\subsubsection{Voronoi Deformation Density}

The Voronoi Deformation Density (VDD) was introduced by Guerra and co-workers in 2004.\cite{bickelhaupt1996,guerra2004} Similar to the Bader method, VDD integrates the electron density over atomic domains to calculate the charge. In case of VDD, these atomic domains are defined by Voronoi polyhedra.\cite{voronoi1908} Furthermore, VDD utilizes the deformation density which is defined as the density change going from Hirshfeld-like promolecules to the final electron density. The charge of atom A is then obtained via integration:
\begin{equation}
    Q_\text{A}^\text{VDD} = - \int\limits_{\text{Voronoi cell}} \left[ \rho(r) - \rho^\text{pro}(r) \right] dr.
\end{equation}
Results for VDD and Hirshfeld are often similar due to their shared methodological roots. Guerra \textit{et al.} benchmarked VDD against Hirshfeld, Bader, and NPA, noting that Bader or NPA overestimate charges.\cite{guerra2004} A recent benchmark study by Choudhuri \textit{et al.} suggest that Hirshfeld analysis tends to underestimate charges which likely affects VDD as well.\cite{choudhuri2020} Similar to AIM methods, VDD results are consistent against the basis set size.

\subsubsection{Charge Model 5}

Charge Model 5 (CM5) was developed by Marenich and co-workers in 2012\cite{marenich2012} and constitutes the latest iteration in the CM\textit{x} (\textit{x} = 1--5 and 4M) series of methods.\cite{storer1995,li1998,winget2002,brom2003,kelly2005,olson2007} CM5 is a semiempirical method that involves fitting charges pre-calculated using the Hirshfeld method with parameters from a test set of dipole moments and charges to overcome the limitations of the base method as discussed in section \ref{s:hirshfeld}. CM5 was benchmarked with promising results.\cite{marenich2012,vilseck2014,choudhuri2020} The CM5 method suffers, like any other (semi-) empirical method, from limited transferability to systems outside the training set.

\subsubsection{Density Derived Electrostatic and Chemical Analysis}

Density derived electrostatic and chemical analysis (DDEC) has been developed over the last decade by Manz and co-workers, with the latest iteration of the approach being DDEC6.\cite{manz2010,manz2012,manz2016,limas2016,manz2017,limas2018} The DDEC approach relies on separation of the electronic density in the vein of the VDD and AIM methods. DDEC6 was tested on pure metals, ice crystals, a large protein system,  and water and ozone molecules, as well as a set of heteroatom-encapsulated endohedral \ce{C60} complexes. In a recent benchmark, Choudhuri and Truhlar compared DDEC5 with their CM5 method as well as with Bader and Hirshfeld charges for various metal dioxides, sulfides, selenides, and metal complex crystal structures.\cite{choudhuri2020} They point out that for the same metal center, DDEC6 charges tend to be systematically bigger or smaller than CM5 charges depending on the type of ligand. For example, DDEC6 charges were found to be 20--25~\% bigger than CM5 charges for \ce{MO2}, \ce{MCO3}, \ce{M(CN)6}, and \ce{M(NCNH)2} structures but smaller by about the same amount for $\ce{MS2}$ and $\ce{MSe2}$ structures. They hypothesize that this behavior may be due to sensitivity of DDEC6 towards the diffuseness of the charge density on the metal centers and advocate for the use of partial charge methods that are validated against dipole moments.

\subsubsection{Charge Equalization Methods} \label{s:eem}

All partial charge analysis methods introduced so far require electronic structure calculations to obtain an optimized charge density of the system under investigation. Although electronic structure calculations can be performed for surprisingly large systems using highly optimized and scalable DFT codes, it is still more common to treat NP model systems using force field methods. In the following, two methods are presented that are constructed to work on the basis of classical potential terms. 

The electronegativity equalization method (EEM) was first introduced by Mortier \textit{et al.} (1985)\cite{mortier_electronegativity_1985,mortier_electronegativity-equalization_1986,van_genechten_intrinsic_1987} based on Sanderson's principle of electronegativity equalization states.\cite{Sanderson1951} With its easily adaptable equations and transferable parameters, the model provides an efficient analysis of charge distributions at a low computational cost. EEM has been validated and successfully applied for applications including organic molecules,\cite{EEM_oc_1,EEM_oc_2,EEM_oc_3,EEM_oc_4,EEM_oc_5,EEM_oc_6,EEM_oc_7,EEM_oc_8,EEM_oc_9} inorganic solids,\cite{EEM_ioc_1,EEM_ioc_2,EEM_ioc_3,EEM_ioc_4,EEM_ioc_5,EEM_ioc_6,EEM_ioc_7} biomolecular systems,\cite{EEM_BioMol_1,EEM_BioMol_2,EEM_BioMol_3,EEM_BioMol_4,EEM_BioMol_5,EEM_BioMol_6} metal-organic frameworks,\cite{EEM_MOF2011,EEM_MOF2012,EEM_MOF2013} and electrochemical systems.\cite{EEM_ec_1,EEM_ec_2,EEM_ec_3} Further improvements and formalisms evolved from the original EEM principle. Derivative methods include the atom-bond electronegativity equalization (ABEEM),\cite{ABEEM_1997} charge equilibration (QEq and CHEQ),\cite{QEQ_1991,CHEQ_2010} fluctuating charges (FlucQ),\cite{flucq_1994} chemical potential equalization (CPE)\cite{CPE_1996}, and split charge equilibration (SQE) approaches.\cite{sqe_2006}

\begin{figure}[htbp]
	\centering
	\includegraphics[width=\linewidth]{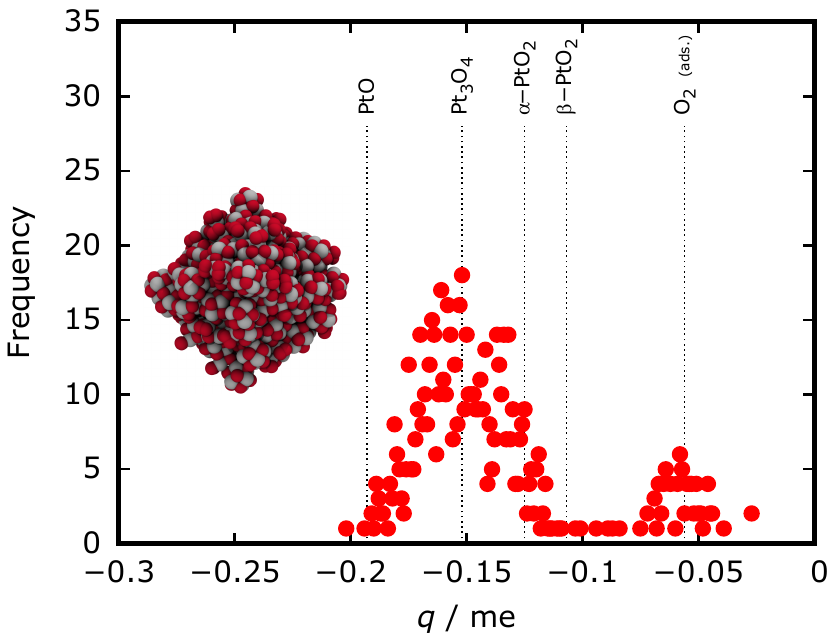}
	\caption{Distribution of EEM partial charges on oxygen atoms for a fully oxidized ($p_\text{O} = 1$~mbar, 500~K) 3 nm octahedral Pt NP obtained using the ReaxFF-GCMC simulation approach. Dotted reference lines indicate the oxygen partial charge distribution maxima obtained from spherical 3~nm reference particles cut from bulk oxide materials. The analysis indicates that the oxide composition of the particle is most closely related to a \ce{Pt3O4} reference particle. Adapted from Kirchhoff \textit{et al.}\cite{kirchhoff2019}}
	\label{fgr:charge_application}
\end{figure}

Our group recently used EEM partial charge analysis to characterize oxidized cuboctahedral platinum NPs with to identify the prevalent type of oxide in this system which at first glance appears amorphous.\cite{kirchhoff2019} An exemplary EEM partial charge distribution is shown in Fig. \ref{fgr:charge_application}.
To provide context for the obtained partial charge distribution, 3~nm spherical particles cut from bulk PtO, $\alpha$-\ce{PtO2}, $\beta$-\ce{PtO2}, and \ce{Pt3O4} are analyzed and the maxima of the oxygen partial charge distributions are shown as dotted lines in Fig. \ref{fgr:charge_application}. Comparison to the reference particles indicates that the oxygen atoms in the oxidized octahedral NP most likely correspond to \ce{Pt3O4}.

Notably, EEM-based approaches suffer from cubic scaling of the dipole polarizability with the system size\cite{EEM_scaling_1,EEM_scaling_2} and often predict fractional molecular charges for systems with well-separated molecules.\cite{EEM_ChargeFlow} Fractional charges result in a conductor-like charge distribution throughout the system and an ill-defined description of oxidized and reduced species.

Verstraelen \textit{et al.} addressed these limitations by proposing the atom-condensed Kohn-Sham density functional theory approximated to second-order (ACKS2) scheme.\cite{acks2_2013,acks2_2014} Here, the EEM formalism has been expanded by additional variables and quadratic energy terms to overcome the unrealistic long-range charge smearing. ACKS2 provides a more realistic description of charge delocalization. 
%
All parameters of the ACKS2 model have direct physical meaning and can be obtained from DFT calculations.\cite{acks2_2013,acks2_2019}

Recent applications of the ACKS2 approach, among others, are the pseudoclassical treatment of explicit electrons in reactive force field simulations,\cite{eReaxFF_1,eReaxFF_2,LiBat_3} proton-transfer reactions in subcritical and supercritical water systems,\cite{manzano_benchmark_2018} and charge-transfer processes in electrolytes of Li-ion batteries.\cite{LiBat_1,LiBat_2,LiBat_3}


\subsection{Radial Atomic Density Distributions}

\begin{figure}[htbp]
\centering
  \includegraphics[width=\linewidth]{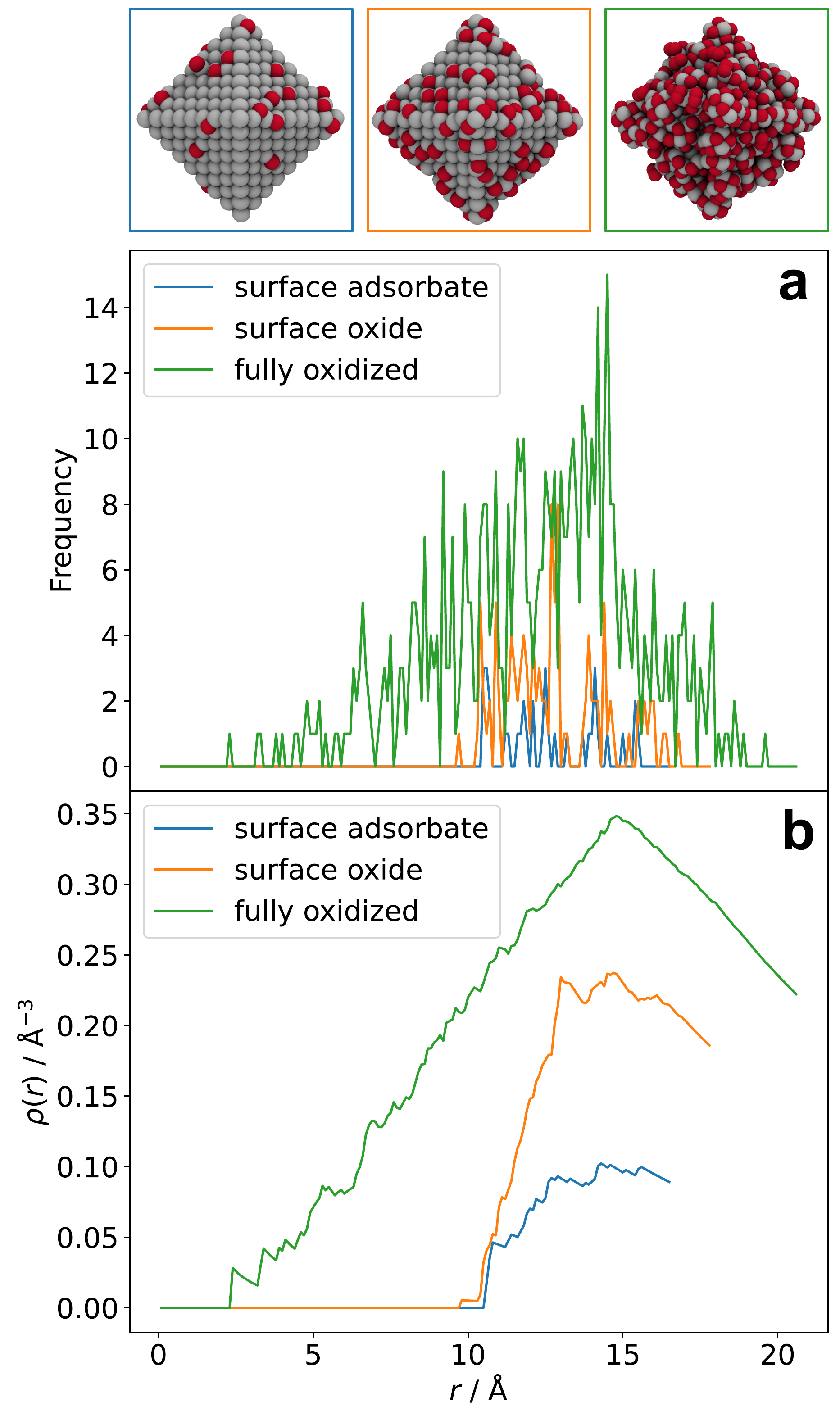}
  \caption{\textbf{a} Radial oxygen distributions of a barely oxidized (blue), surface oxidized (orange), and fully oxidized (green) 3~nm octahedral Pt NPs obtained from ReaxFF GCMC calculations. \textbf{b} Integrated, normalized radial oxygen density distributions of the same structures.}
  \label{fgr:radial_density}
\end{figure}

Radial atomic density distributions visualize the distribution of a particular chemical species in a complex heteroatomic material. Senftle and co-workers used it to illustrate different stages of oxidation of a Pd NP,\cite{senftle2013} hydrogen uptake of a Pd NP\cite{senftle2014-1}, as well as Pd carbide formation.\cite{senftle2014} To obtain such a distribution, atoms of the respective chemical species are counted in volume segments radially outward from the center of the system. In Fig. \ref{fgr:radial_density} \textbf{a}, radial atomic density distributions are used to show the distribution of oxygen atoms in barely oxidized (blue), surface-oxidized (orange), and fully oxidized (green) 3~nm cuboctahedral Pt NPs.
While oxygen atoms are mostly located in the outer shell of the particle at surface oxidation conditions (blue and orange), oxidation all the way to the core can be observed beyond this point (green). Broadening of the green distribution indicates expansion of the NP as a result of oxidation.

Integration over the volume segments and normalization of the results gives the distributions shown in Fig. \ref{fgr:radial_density} \textbf{b}. Compared to Fig. \ref{fgr:radial_density} \textbf{a}, this way of presentation is less noisy and allows for large amounts of data to be presented at once. For example, Fig. \ref{fgr:radial_density} \textbf{b} presents integrated distributions for the three aforementioned barely oxidized (blue), surface-oxidized (orange), and fully oxidized (green) structures. While Fig. \ref{fgr:radial_density} \textbf{b} still retains information about the oxidation-dependent expansion of the particles, the regular radial atomic density distribution in Fig. \ref{fgr:radial_density} \textbf{a} can be easier to read depending on what information is most important to visualize; for example, it is immediately obvious from the absolute Frequency value in Fig. \ref{fgr:radial_density} \textbf{a} that the highest density of O atoms in the configuration obtained at 400~K is found between 12 and 14~\AA, whereas this information is contained in the slope of the corresponding data in Fig. \ref{fgr:radial_density} \textbf{b}.

Code examples for calculating (integrated) radial atomic density distributions from an atomic configuration can be found in the supporting git repository in \verb|densityDist.py|.

\subsection{Oxidation State Analysis}

\begin{figure}[htbp]
    \centering
    \includegraphics[width=\linewidth]{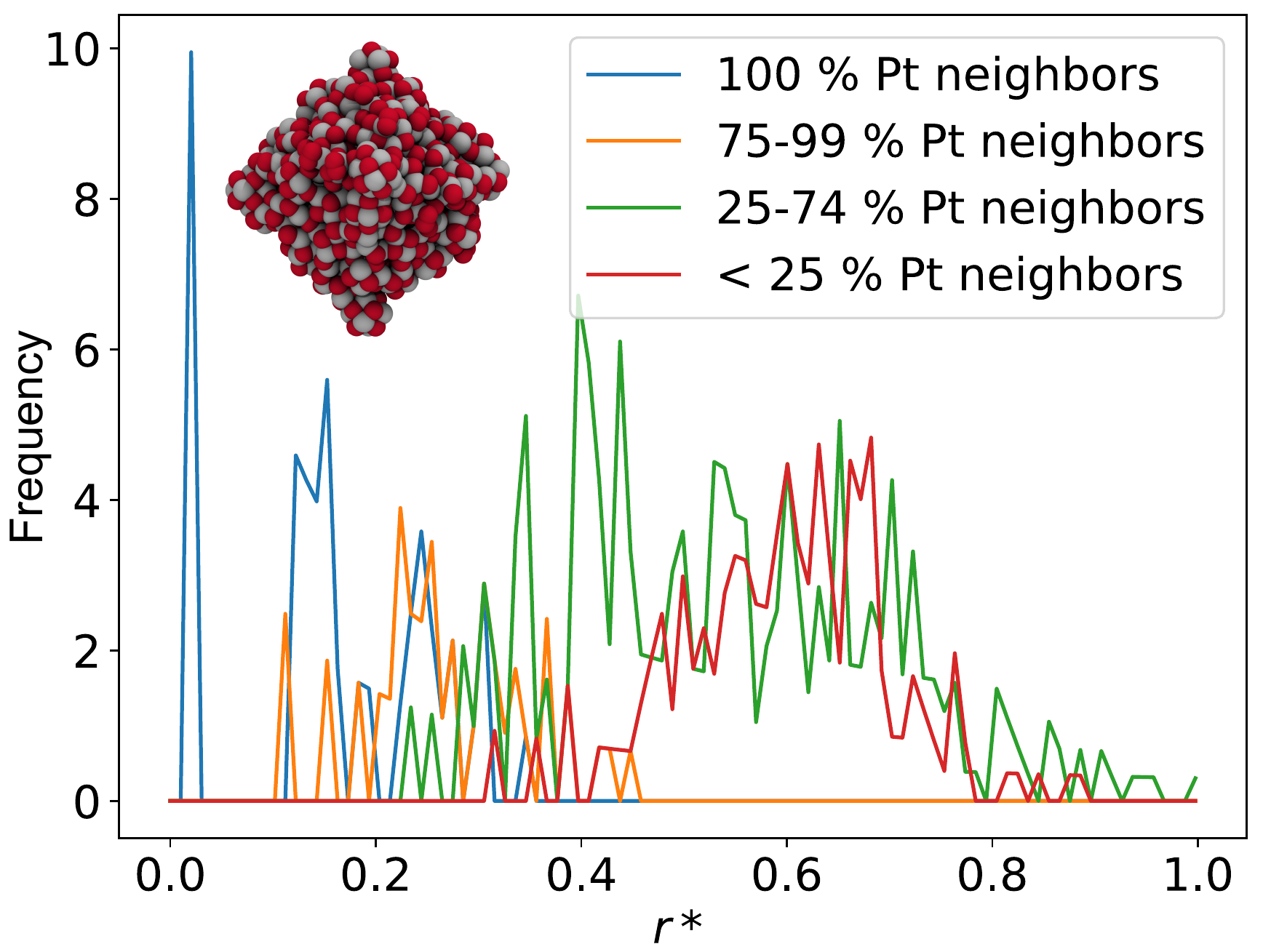}
    \caption{A simplified oxidation state analysis of a strongly oxidized 3~nm octahedral Pt NP. The analysis shows that the particle has some core-shell character, with a strongly oxidized exterior (0--74~\% Pt neighbors, red and green) but some remaining metallic states at the center (100~\% Pt neighbors, blue). Different oxidation states of Pt atoms are estimated by the ratio between the total number of nearest-neighbor atoms and the number of nearest-neighbor oxygen atoms. Note that in this evaluation, undercoordinated Pt atoms at edges and kink sites with no oxygen neighbors still count as fully metallic (100~\% Pt neighbors). Depending on the desired information, the ratio may also be calculated with respect to the maximum number of bonds in the bulk crystal to account for the unsaturated nature of edge and kink sites.}
    \label{fgr:oxidationstate}
\end{figure}

The oxidation state of an atom in a metallic NP can be estimated by the number of its metal-metal bonds. A density of oxidation states can thus be obtained by showing the number of atoms within a specific oxidation state (\textit{i.e.} number of metal-metal bonds) as a function of the radial distance \textit{r} from the particle center. Hong and van Duin\cite{hong2015} used this technique to monitor the formation of different types of oxides during a ReaxFF-driven reactive molecular dynamics (MD) simulation of an aluminium NP in the presence of oxygen. Fig. \ref{fgr:oxidationstate} shows an application of the method to a strongly oxidized 3~nm octahedral Pt NP. 

A code example for performing oxidation state analysis from an atomic configuration can be found in the supporting git repository in \verb|oxidationStates.py|.

\subsection{Radial Distribution Function}

A radial distribution function (RDF, $g(r)$) is obtained by calculating distances between all pairs of atoms in a system and presenting them as a volume-normalized density ($\rho = N / V$). Usually, the density is obtained by counting all distances that fall into $r + dr$ sphere segments originating from the center of the system. The RDF is therefore entirely dependent on the atomic configuration of the system.

The $g(r)$ can be measured in X-ray\cite{sirota1989,soper2007} or neutron scattering experiments.\cite{gingrich1961,yarnell1973,li1990,soper2007} Theory can play an integral role in the analysis of scattering data. For example, hybrid quantum mechanical / molecular mechanical (QM/MM) simulations have been successfully used to predict RDFs and assist in the interpretation of the experimental result.\cite{soper2007,dohn2015,dohn2016}

\begin{figure}[htbp]
    \includegraphics[width=\linewidth]{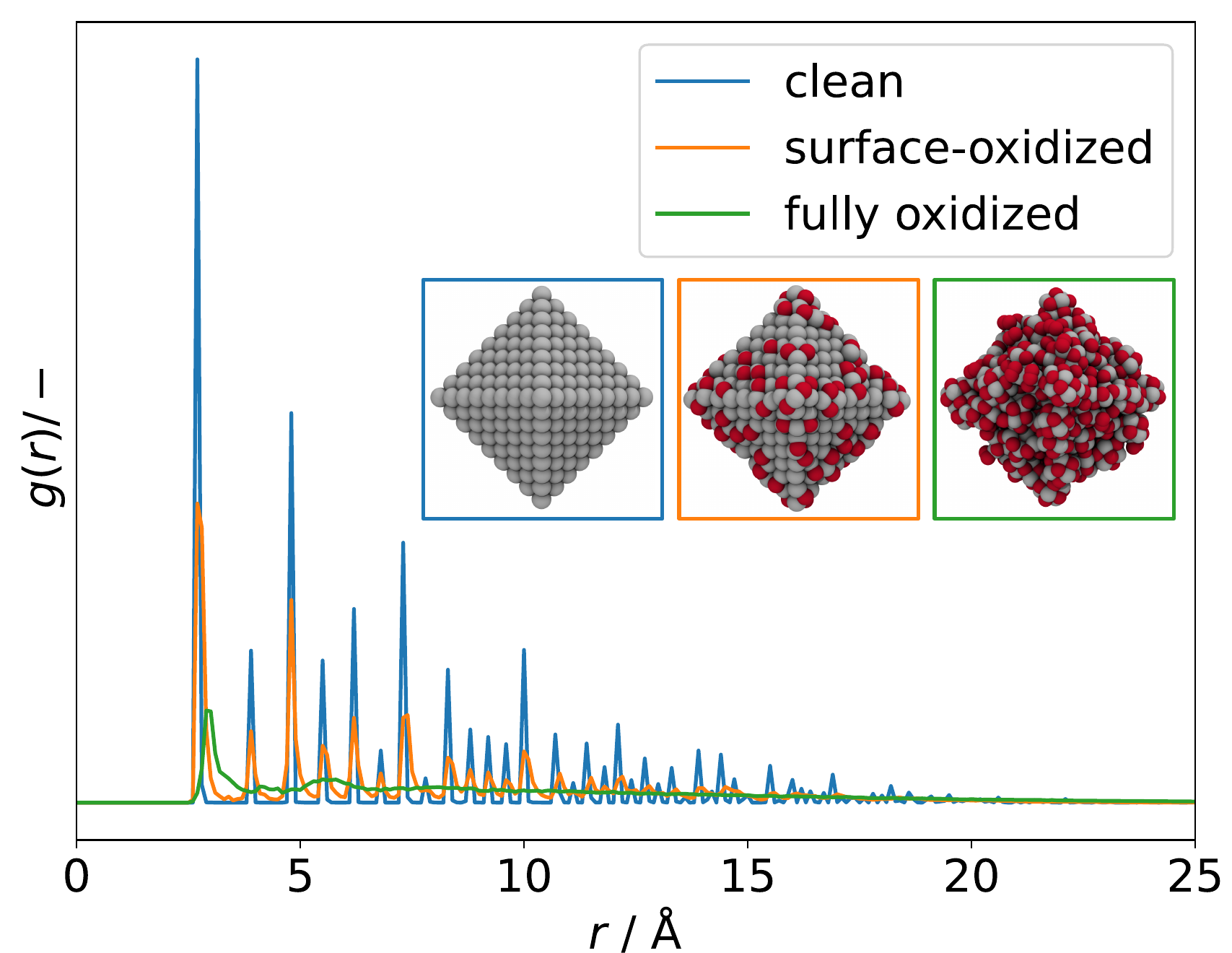}
    \caption{RDFs of clean (blue), surface-oxidized (orange), and fully oxidized (green) 3~nm octahedral Pt NPs obtained from ReaxFF-GCMC simulations. Significant broadening of the peaks is observed as long-range ordering of the particles decreases as a result of progressing oxidation.}
    \label{fgr:rdfs}
\end{figure}

Going back to our guinea pig, the 3~nm octahedral Pt NPs, the $g(r)$ can be used to quantify the degree of amorphicity in a series of increasingly oxidized structures. Fig. \ref{fgr:rdfs} illustrates RDFs of an almost clean octahedral particle (blue) as well as of a surface-oxidized (orange) and of a fully oxidized particle (green).  
The RDFs become more and more liquid-like as oxidation proceeds and long-range ordering is lost. 

A recent example of RDFs applied to NP systems was published by Thom{\"a} \textit{et al.} who experimentally measured the RDF response of the water solvation shell around faceted iron oxide NPs.\cite{thoma2019} The group used calculated RDFs of different bonding situations of water to assign the correct structures to the measured bands. In another recent computational study, Zeng \textit{et al.}\ used RDFs to monitor the structure and response to heating of a core-shell aluminium / aluminium oxide NP and its surrounding water shell over the course of a ReaxFF reactive MD simulation.\cite{zeng2018}

A code example for calculating RDFs from an atomic configuration can be found in the supporting git repository in \verb|RDF.py|.

\subsection{Common Neighbor Analysis}

Common neighbor analysis (CNA) was introduced by Honeycutt and Andersen in 1987.\cite{honeycutt1987} CNA seeks to identify reoccurring structural patterns in atomic configurations. CNA is based on earlier works by Blaisten-Barojas\cite{blaisten1984} and Haymet\cite{haymet1984} and was further developed and first implemented into code by Clarke, Faken, and J{\'o}nsson.\cite{clarke1993,faken1994} In CNA, each pair of atoms in a structure is assigned a triple of indices, \textit{jkl}.\footnote{Note that use of quadruples is proposed in the original publication by Honeycutt and Andersen.\cite{honeycutt1987} We focus on the more commonly used triples representation introduced by Clarke and J{\'o}nsson in 1993.\cite{clarke1993}} The three indices represent the number of nearest neighbors common to both atoms (\textit{j}), the number of bonds between common neighbors (\textit{k}), and the number of bonds in the longest chain formed by the \textit{k} bonds between common neighbors (\textit{l}). These triples are unique for characteristic structural motifs such as FCC, HCP, BCC, or icosahedral arrangements of atoms. Large structures can therefore be characterized based on the occurrence of smaller sub-structures with fingerprint triples. While relatively insensitive to small displacements of atoms, the method will quickly highlight larger changes in crystal structure caused, for example, by local melting, and can therefore be used to analyze phase transitions. In their original publication, Honeycutt and Andersen annealed small Lennard-Jones clusters of 13 to 309 atoms in MD simulations and used CNA to identify the dominant structural feature (FCC, HCP, or icosahedral) of the relaxed structures. 

CNA can be understood as a partitioning of the RDF of the system. The RDF, $g(r)$, of a structure can be represented as a sum of RDFs of each \textit{jkl} triple, $g_{jkl}(r)$, so that
\begin{equation}
    g(r)  = \sum\limits_{jkl} g_{jkl}(r). \label{gjkl}
\end{equation}
This renders $g_{jkl}(r)$ an interesting tool for the interpretation of the whole $g(r)$. For example, by integrating the first peak in $g_{jkl}(r)$, the number of pairs $N_{jkl}$ with \textit{jkl} can be determined as
\begin{equation}
    N_{jkl} = \frac{4 \pi N}{V} \int\limits_{0}^{r_c} r^2  g_{jkl}(r) \text{d}r. \label{Njkl}
\end{equation}
$N$ is the total number of atoms in the system and $r_c$ is the integration cutoff. By choosing integration limits accordingly, other peaks can be analyzed as well. The method is sensitive to changes to $r_c$ since the cutoff radius determines which atoms are considered nearest-neighbors and, therefore, which triples are found.

As exemplary applications, CNA has been used to describe solid-liquid phase transitions such as melting of small clusters,\cite{cleveland1998,cleveland1999} strain and deformation in metallic systems,\cite{sorensen1998} and properties of glasses.\cite{cheng2011} CNA routines can be found in many computational tools, such as the open-source Atomic Simulation Environment (ASE).\cite{larsen2017}

A code example for calculating CNA from an atomic configuration can be found in the supporting git repository in \verb|CNA.py|.

\subsection{X-Ray Diffraction Spectrum}

(Powder) X-ray diffraction (XRD) spectra of NPs can be calculated from a given structure using the Debye scattering equation
\begin{eqnarray}
    I(q) &=& \sum\limits_{i=1}^{N}\sum\limits_{j=1}^{N}\ f_i(q)\ f_j(q)\ \frac{\sin(q \times r_{ij})}{q \times r_{ij}}\ \ \text{with}\\
    q &=& \frac{\sin \theta}{\lambda},
\end{eqnarray}
where $I$ is the resulting scattering intensity, $q$ is the scattering vector with the incidence angle $\theta$ and the incidence wavelength $\lambda$, $i$ and $j$ are atomic indices, $N$ is the total number of atoms in the system, $f_i$ and $f_j$ are scattering factors specific to the chemical species of atoms $i$ and $j$, and $r_{ij}$ is the distance between atoms $i$ and $j$. The scattering factors $f$ can be calculated from tabulated values\cite{cromer1968,brown2006} using the relationship
\begin{equation}
    f(q) = \sum\limits_{i=1}^4 a_i\ \exp \left(-b_i\ \frac{\sin^2\theta}{\lambda^{2}}\right) + c,
\end{equation}
where $a_i$,  $b_i$, and $c$ are the tabulated parameters ($i = 1-4$), $\theta$ is the incidence angle and $\lambda$ is the incidence wave length.

\begin{figure}[htbp]
    \centering
    \includegraphics[width=\linewidth]{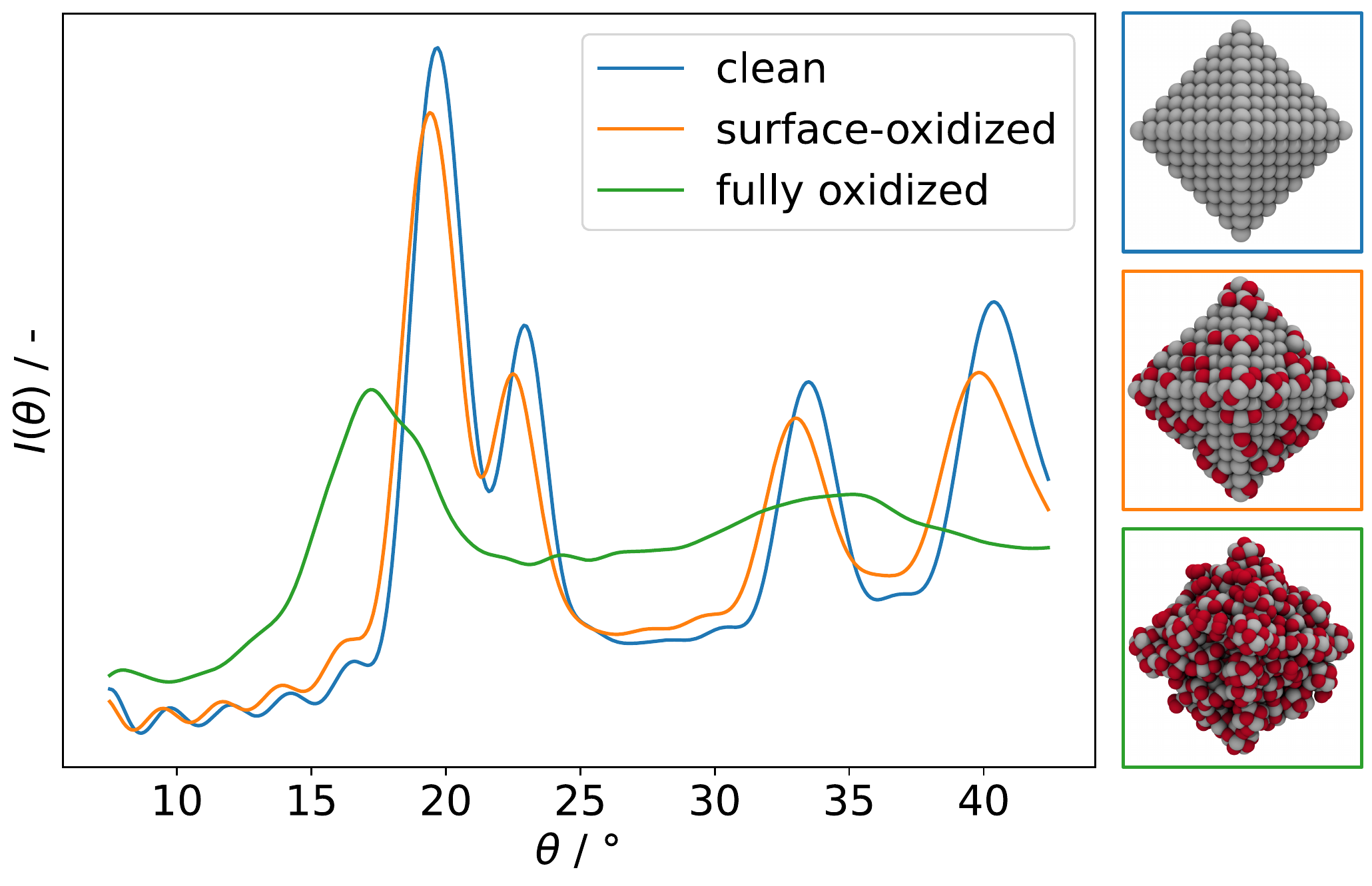}
    \caption{Simulated powder XRD spectra of clean (blue), surface-oxidized (orange), and fully oxidized (green) 3~nm octahedral Pt NPs. Assuming Cu-\textit{K}$_\alpha$ incidence radiation.}
    \label{fgr:xrds}
\end{figure}

Fig. \ref{fgr:xrds} compares powder XRD spectra of barely oxidized (blue), surface oxidized (orange), and fully oxidized (green) 3~nm octahedral Pt NPs.
Band broadening is observed already for the barely oxidized system as a result of its nanometer size.\cite{holder_tutorial_2019} As oxidation proceeds, bands broaden even further and the overall intensity $I$ reduces, indicating that the NP becomes increasingly disordered. Similar to RDFs, with which this approach has in common the dependence on atomic pair distances, the calculated powder XRD spectra can therefore be used to quantify the degree of amorphicity in a series of structures aside from determining the characteristic diffraction peaks.

Examples for successful use of simulated XRD spectra in literature are abundant. To name a few, Chiche and co-workers showed that calculated XRD spectra can be used in combination with measurements to reveal detailed structural information.\cite{chiche2008} By comparing simulated and experimental XRD spectra of MgO and bohemite, the group showed that the simulated spectra contain detailed information about size and shape of the studied 3--7~nm NPs. They note that XRD in such cases outperforms TEM measurements which fail to provide morphological details for particles in this size range. Naicker \textit{et al.} used simulated XRD spectra to monitor the characteristic rutile, anatase, and brookite bands during MD simulations of \ce{TiO2} NPs to study surface rearrangements.\cite{naicker2005} Vogel and co-workers used simulated and measured powder XRD spectra to characterize the reversible uptake of H in Pd NPs.\cite{vogel2010} Similarly, Gilbert and co-workers used simulated powder XRD spectra in combination with extended X-ray absorption fine structure (EXAFS) and wide-angle X-ray scattering (WAXS) spectroscopy techniques to study the increase in crystallinity in ZnS NPs upon hydration.\cite{gilbert2004}

A code example for calculating powder XRD patterns from an atomic configuration using the Debye formalism can be found in the supporting git repository in \verb|XRD.py|.

\subsection{Solvent-Accessible Surface Area and Interaction Heatmaps}

The Solvent-Accessible Surface Area (SASA) method can be used to create heatmaps of the interaction strength of probe atoms or molecules with a nanoparticle structure. The SASA method is rooted in biomolecular applications and computational protein structure prediction\cite{DurhamSASA, LeeSASA} and was originally proposed by Richmond in 1984.\cite{richmond1984} To calculate the SASA, the van der Waals (vdW) radius of each surface atom in a structure is extended by a certain distance, for example by the vdW radius of the probe.\cite{LeeSASA} By systematically placing probe atoms (\textit{e.g.} oxygen, hydrogen, or a water molecule) on the extended vdW surface, an interaction heatmap visualizing the strength of interaction of the structure with the probe is obtained. 

\begin{figure}[htbp]
    \centering
    \includegraphics[width=\linewidth]{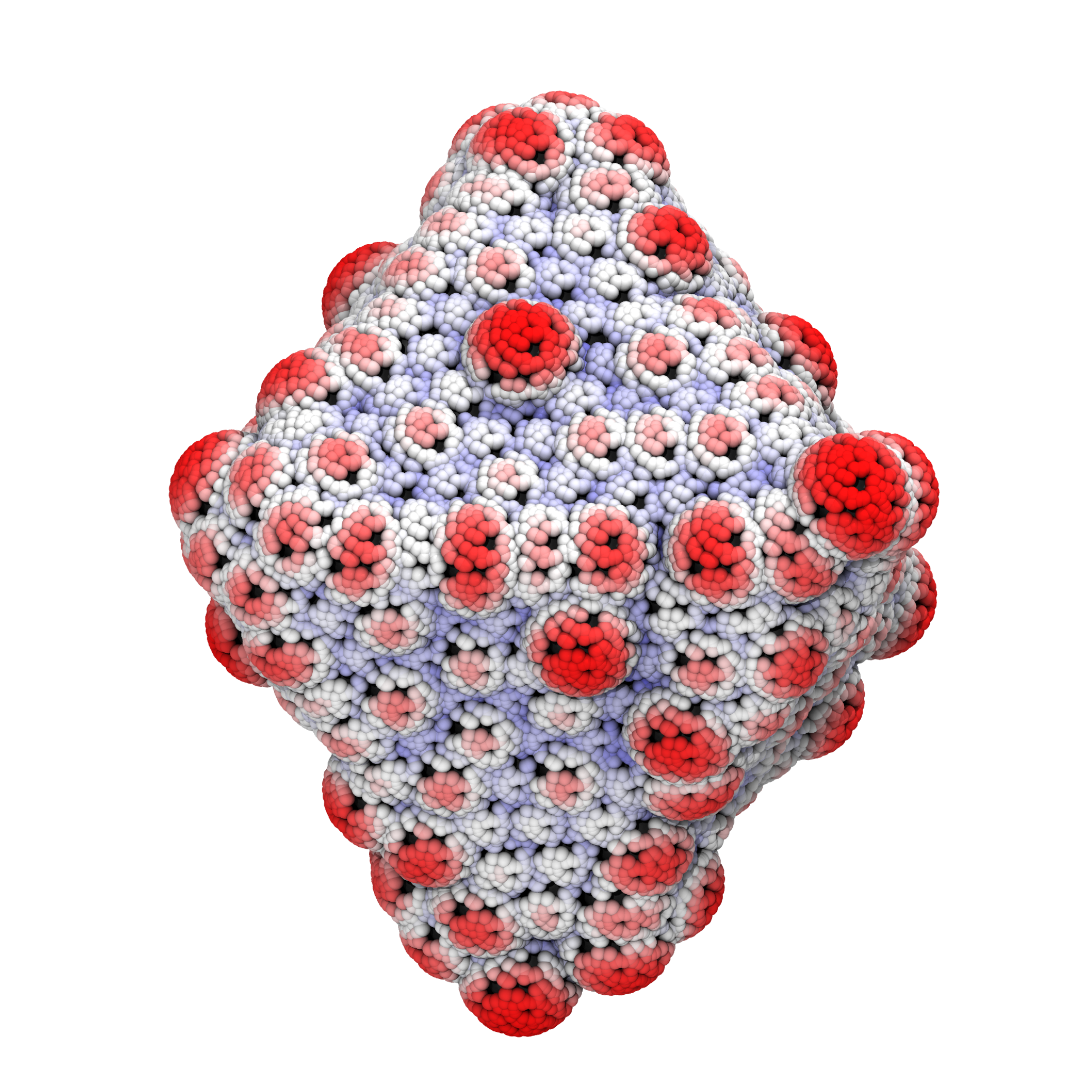}
    \caption{Interaction heatmap for an oxygen probe atom with a fully oxidized 3~nm octahedral Pt NP. Equidistant interaction points on the expanded vdW surface of the NP are used. Red: strong interaction; white: weak interaction.}
    \label{fig:heatmap}
\end{figure}

Fig. \ref{fig:heatmap} shows the interaction heatmap for an oxygen probe atom with an oxidized 3~nm octahedral Pt NP.

\begin{figure}[htbp]
    \centering
    \includegraphics[width=0.75\linewidth]{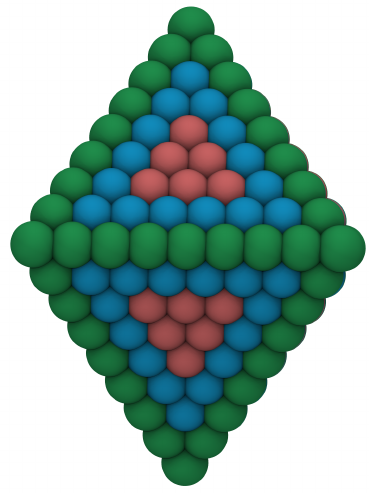}
    \caption{Interaction heatmap of an oxygen probe atom with a pristine 3~nm octahedral NP. Interaction strengths for the surface atoms are calculated by averaging over interaction points on the SASA nearest to each surface atom. Interaction strength: green > blue > red.}
    \label{fig:heatmap_sum}
\end{figure}

By assigning interaction points to the respective nearest surface atoms and averaging the interaction energy results assigned to each atom, the interaction strength of each surface atom can be estimated. This approach is exemplified in Fig. \ref{fig:heatmap_sum} for a pristine 3~nm octahedral Pt NP, using an oxygen probe atom. 
In this specific case, the NP edges are revealed to interact most strongly with the oxygen probe, suggesting that adsorption on these sites is preferred during early stages of the oxidation.\cite{kirchhoff2019,braunwarth_exploring_2020} The SASA heatmap approach therefore constitutes an intuitive way of visualizing surface reactivity even for complex systems.

\subsection{Two-Phase Thermodynamics}

The Two-Phase Thermodynamics (2PT) method enables the calculation of the density of states (DoS) of liquids from molecular dynamics trajectories, thus allowing the identification of translational, rotational and intramolecular vibrational modes.\cite{Lin032PT,Lin102PT,Pascal2PT} The 2PT method can aid in the interpretation of complex data obtained using surface infrared or Raman spectroscopy methods, in particular for studying the solid-electrolyte interface.\cite{Heyden2PT,Goddard2PT}

In the 2PT scheme, the velocity autocorrelation function (VACF) is calculated via the time-dependent atomic velocities, giving access to the DoS function. The total DoS is then separated into a diffusive gas-like component and a solid-like component. The components are treated separately via hard sphere thermodynamics and quantum statistics, enabling calculation of entropy and heat capacity for example. 

The 2PT scheme has been used successfully in literature. In a reactive MD based study, Cheng \textit{et al.} identified intermediates of the \ce{CO2} reduction reaction by their characteristic bands in the vibrational DoS.\cite{Goddard2PT} The group included explicit solvation and the electrode potential in their simulations. Pascal and co-workers evaluated the translational, rotational and vibrational entropic contributions of confined water using the 2PT method from MD simulations. They studied the wetting behavior of hydrophilic carbon nanotubes and report that water in the nanotube is found to be more stable despite the confinement.\cite{Yousung2PT} Persson and co-workers introduced a spatially resolved 3D variant of the 2PT method.\cite{Heyden2PT} Using MD simulations, the group studied local contributions to the solvation enthalpy, entropy, and free energy of hydrophilic and hydrophobic small molecular solutes under the influence of ions and the hydrogen bonding network.

\section{Analyzing Sets of Calculations}

\subsection{Normalized Formation Energy}

\begin{figure}[htbp]
    \centering
    \includegraphics[width=\linewidth]{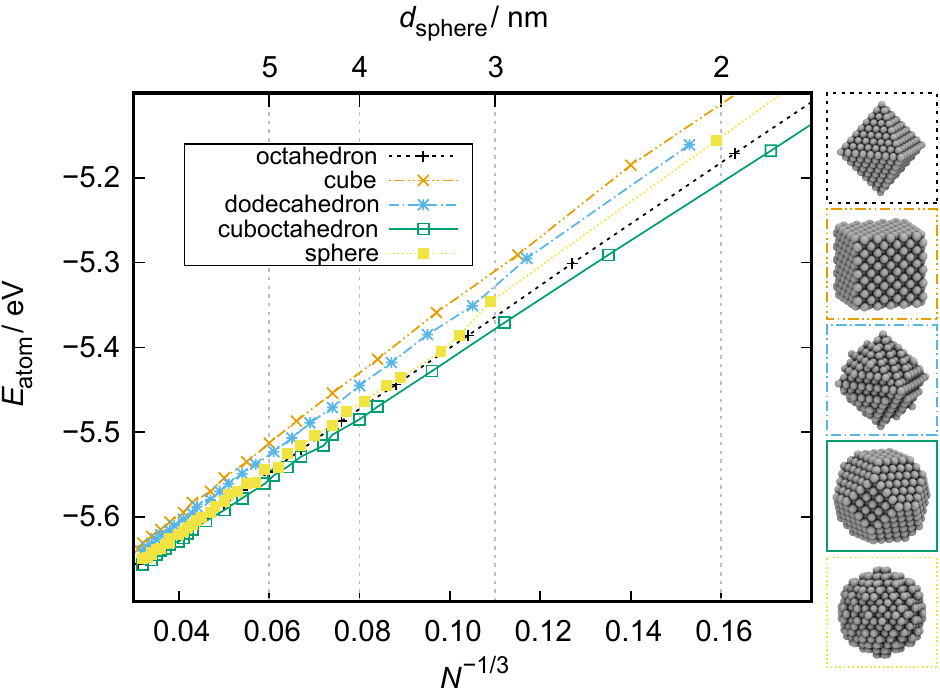}
    \caption{Stability of cuboctahedral, octahedral, spherical, dodecahedral, and cubic Pt NPs as a function of particle size obtained from ReaxFF calculations. $N$: number of atoms in a particle. The key diameter of the spherical particle, $d_\text{sphere}$, is given on top. Bottom: illustration of NP structures of \textit{ca.} 3 nm size. Adapted from Kirchhoff \textit{et al.}\cite{kirchhoff2019}}
    \label{fig:formationenergy}
\end{figure}

To study the stability of different NP structures as a function of particle size, the total energy per atom, $E_\text{atom}$, can be set in relation with the the NP size to obtain stability trends. Such analysis was performed, for example, by Huang \textit{et al.} on Pt NPs of various shapes from 20 to 100 nm.\cite{huang2011} Another useful variant of this analysis is to show $E_\text{atom}$ as a function of the cubic root of the number of atoms in the system, $N^{-1/3}$. Kirchhoff \textit{et al.} used this type of analysis to study the stability of cuboctahedral, octahedral, spherical, dodecahedral, and cubic Pt NPs in a size range of \textit{ca.}~1--10~nm.\cite{kirchhoff2019} The results are reproduced in Fig. \ref{fig:formationenergy}.
Displaying the energy per atom as a function of $N^{-1/3}$ linearizes the data. This allows for intuitive comparison of the stability of, in this example, different NP shapes. Differences between the particles are most distinct at small sizes and values converge to the bulk $E_\text{atom}$ for the ReaxFF force field that they were obtained with, which is 5.77~eV atom$^{-1}$.\cite{fantauzzi2014} This linearized presentation also reveals slight alterations in the facet composition. For example, small differences in the relative size of (111) and (100) facets of cuboctahedral particles show as deviations from the otherwise linear behavior since the surface-to-volume ratio is not constant anymore. The same observation can be made for slightly trunctated octahedral or dodecahedral particles or when the edges of the cubic particle start to round off, for example as a result of surface reactions that take place.

\subsection{Adsorption Isotherms and Isobars} \label{s:isobars}

Adsorption isobars and isotherms can reveal information about the structure of a system in the presence of an adspecies at different thermodynamic conditions. Adsorption isobars or isotherms are constructed by showing the surface coverage as a function of either the partial pressure of the adspecies or the system temperature while keeping constant the other respective other value. For example, Senftle \textit{et al.} constructed hydrogen adsorption isotherms to explore the uptake of hydrogen in Pd as a function of Pd cluster size.\cite{senftle2014-1}

A data set for this type of analysis can be achieved in different ways. One route is to manually generate structures of increasing coverage and diverse arrangements of an adspecies $x$ and to calculate the thermodynamic stability of each structure at various different chemical potentials $\mu_x(p,T)$ in order to find the most stable coverage for each condition. Alternatively, schemes like the grand-canonical Monte Carlo (GCMC) algorithm\cite{senftle2013,senftle2014,senftle2014-1} can be used to generate diverse data sets of structures at certain $\mu(p,T)$ conditions using a stochastic sampling approach. 

\begin{figure}[htbp]
    \centering
    \includegraphics[width=\linewidth]{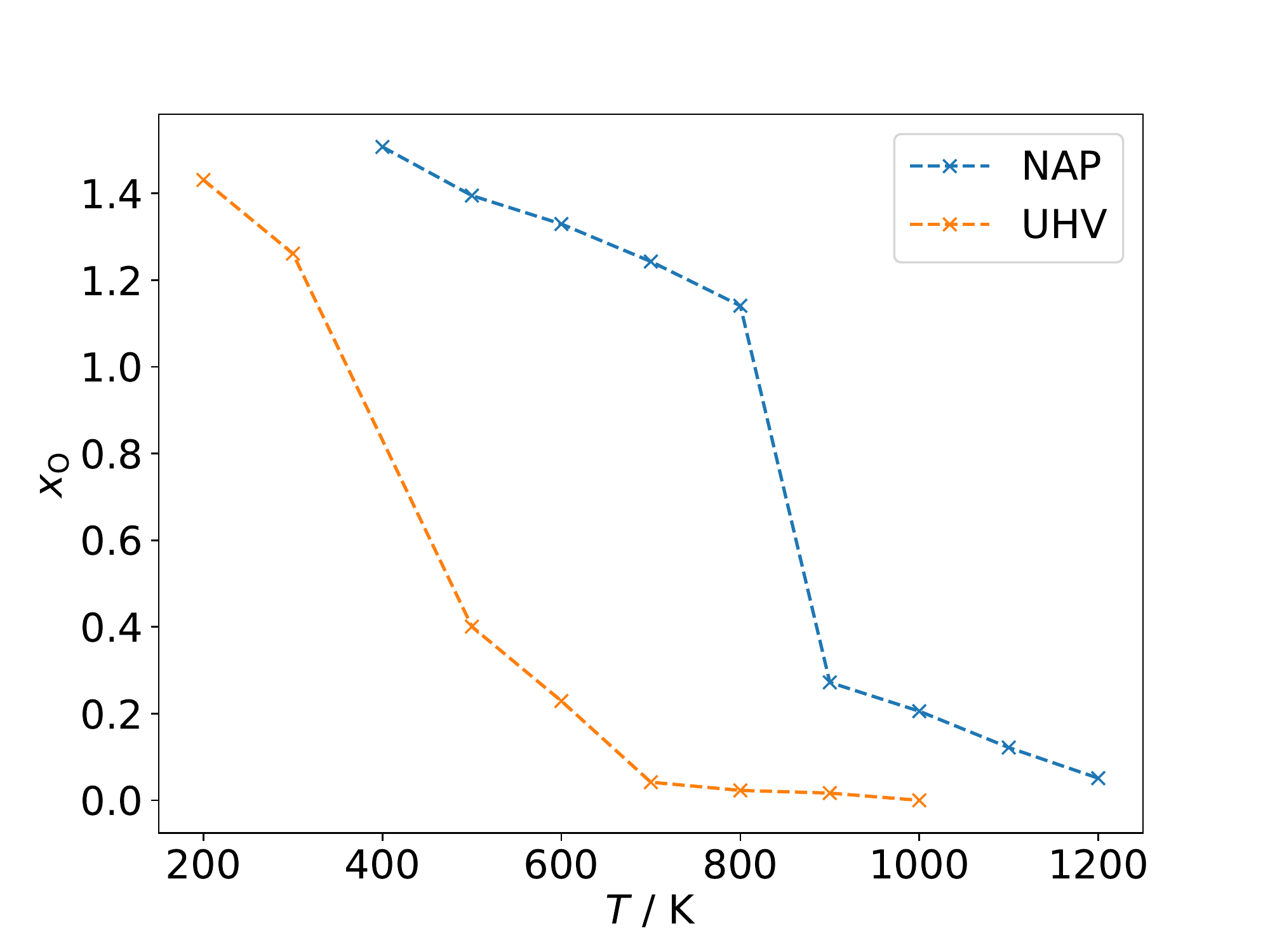}
    \caption{Adsorption isobars for a 3~nm octahedral Pt particles oxidized at UHV ($p_\text{O} = 10^{-10}$~mbar) and NAP ($p_\text{O} = 1$~mbar) conditions and 200--1200~K temperature. The degree of oxidation is quantified as the quotient between oxygen and platinum atoms in the system ($x_\text{O} = N_\text{O}$ / $N_\text{Pt}$). Adapted from Kirchhoff \textit{et al.}\cite{Kirchhoff2022}}
    \label{fig:isobars}
\end{figure}

Returning to the data set of oxidized 3~nm octahedral Pt particles, which was generated using a reactive force field and the GCMC scheme at oxygen chemical potentials $\mu_\text{O}(p,T)$ corresponding to 200--1200~K at UHV ($p_\text{O} = 10^{-10}$~mbar) and NAP ($p_\text{O} = 1$~mbar),\cite{Kirchhoff2022} adsorption isobars can be constructed as shown in Fig. \ref{fig:isobars}.
The coverage convention introduced in section \ref{s:coverage} is used. The individual data points shown in Fig. \ref{fig:isobars} correspond to the structures with the most favorable energy of formation at the corresponding $\mu_\text{O}(p,T)$.

Fig. \ref{fig:isobars} reveals that the oxidation process in case of this particular model system can be broadly divided into three steps: phase 1, surface adsorption at high temperatures marked by a slow increase of the $N_\text{O} / N_\text{Pt}$ ratio; phase 2, quick increase of $N_\text{O} / N_\text{Pt}$ marking full oxidation of the particle; and phase 3, decoration of the oxidized particles by additional dioxygen species on the surface.\cite{kirchhoff2019} This trend is most obvious in case of the more aggressive NAP conditions.

A code example for evaluating such adsorption isobars from a data set of oxidized Pt NPs can be found in the supporting git repository in \verb|adsorptionIsobars.py|.

\subsection{Adsorbate Phase Diagrams}

\begin{figure*}[tb]
    \centering
    \includegraphics[width=\linewidth]{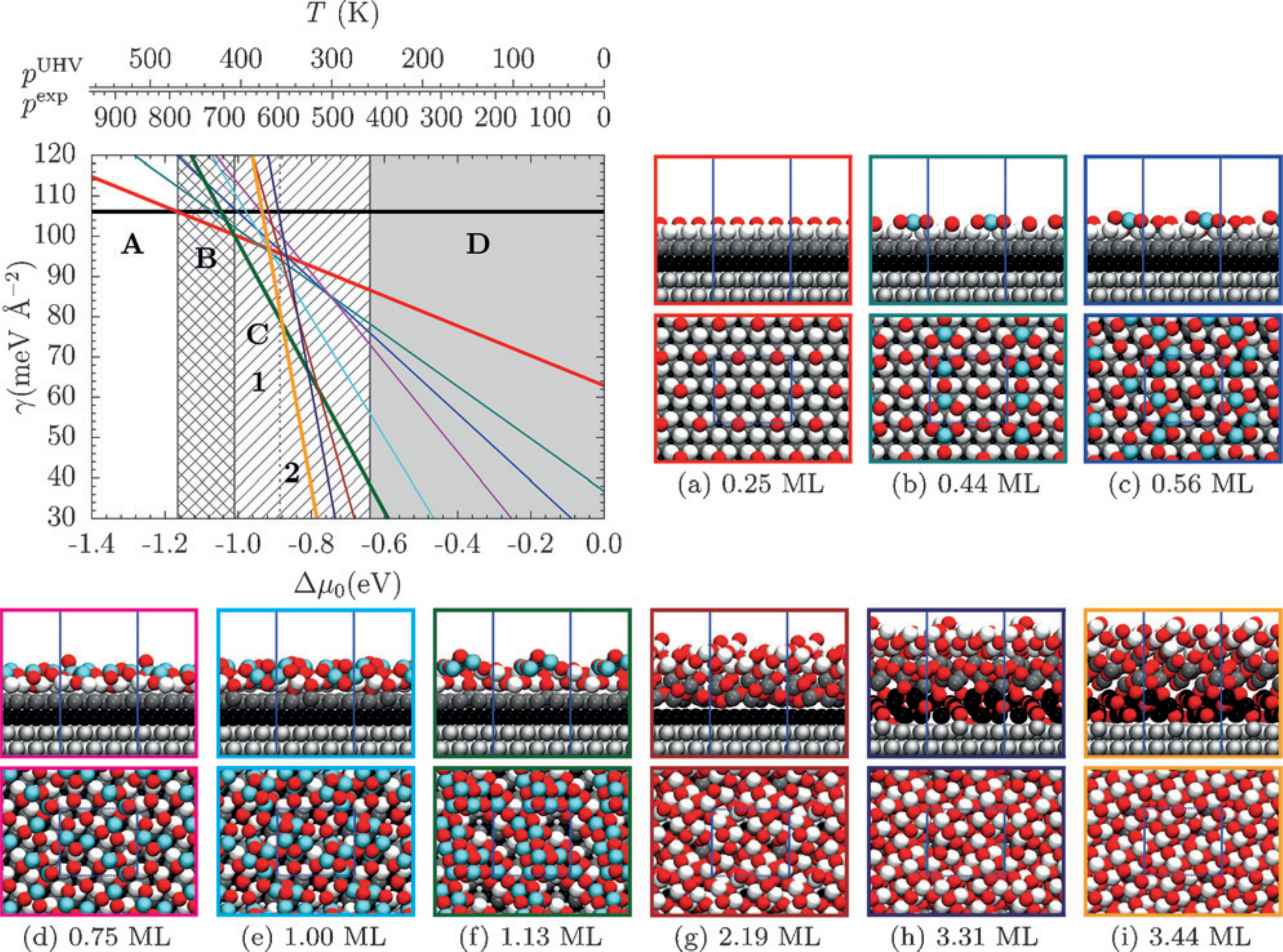}
    \caption{Phase diagram for the adsorption of a Pt(111) model system. The data set was obtained using ReaxFF-GCMC simulations at a MC temperature of 650 and 600~K. The colored boxes (a)–(i) correspond to the lines in the phase diagram. Pt atoms are shown in shades of grey depending on the layer or light blue when buckled (only for $\Theta_\text{O} \leq 1.13$~ML). The phase diagram shows five thermodynamically stable phases: A clean Pt(111), B \textit{p}(2$\times$2) adsorbed atomic oxygen, C a low- and a high-coverage surface oxide phases (1 and 2) and D $\alpha$-\ce{PtO2} bulk oxide. Reprinted with permission from Fantauzzi \textit{et al.}, \textit{Angew. Chem. Int. Ed.}, 2017, \textbf{56}, 2594-2598. Copyright 2017 by John Wiley \& Sons.}
    \label{fig:phasediagram}
\end{figure*}

The thermodynamic adsorption phase diagram can be regarded as a generalized case of an adsorption isotherm and isobar (section \ref{s:isobars}). To construct a phase diagram, the energy of formation is displayed as a function of the chemical potential $\mu(p,T)$ of the adspecies. The formerly separated quantities of temperature and pressure are thus combined in this case. An exemplary phase diagram for adsorption of oxygen on Pt(111) published by Fantauzzi and co-workers\cite{fantauzzi2017} is given in Fig. \ref{fig:phasediagram}.
This type of analysis can be used to predict the approximate coverage under given thermodynamic conditions and can thus be used to supplement experimental assessments\cite{fantauzzi2015} and predict new structures.\cite{fantauzzi2017}

Adsorbate phase diagrams can quickly become complex as the number of variables increases, for example in case of co-adsorption of multiple chemical species. Ferguson and co-workers have therefore developed a visualization tool entitled Surface Phase Explorer (SPE) available through \href{https:\\nrel.gov}{nrel.gov} to facilitate visualization of multidimensional phase diagrams.\cite{ferguson_ab_2016}

\begin{figure}[htbp]
    \centering
    \includegraphics[width=\linewidth]{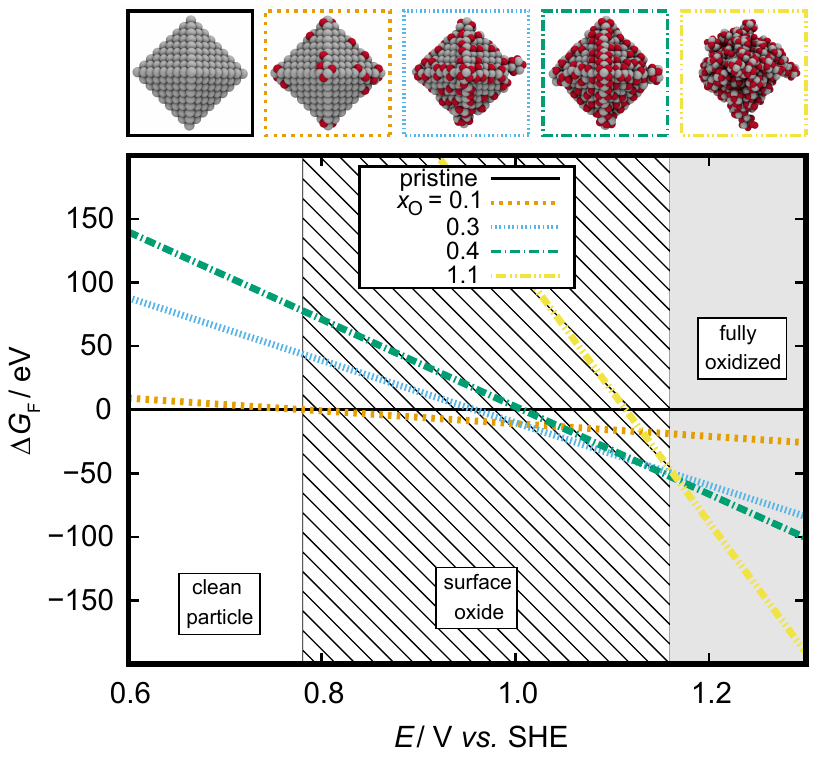}
    \caption{Electrochemical phase diagram depicting the most stable adsorbate structures on a 3~nm octahedral Pt NP. Adsorbate structures are characterized by their O:Pt ratio, $x_\text{O}$, as outlined in section \ref{s:coverage}. Adapted from Kirchhoff \textit{et al.}\cite{kirchhoff2019}}
    \label{fig:echemphasediagram}
\end{figure}

Finally, phase diagrams can be brought into the realm of electrochemistry using the Extended \textit{Ab Initio} Thermodynamics (EAITD) approach.\cite{jacob2007,kaghazchi_bridging_2009,venkatachalam_density_2010} This approach makes use of the computational Standard Hydrogen Electrode and therefore assumes that the reaction \ce{H2 <-> 2 H+ + 2e-} is in equilibrium (assuming pH 0, 298~K, 1~atm pressure).\cite{trasatti_absolute_1986} The SHE can be applied to all electrochemical reactions that involve protons in any way. For our exemplary data set of oxidized 3~nm octahedral Pt NPs, we can construct a thermodynamic cycle based on the assumption that all oxygen atoms must come from the water splitting reaction, \ce{H2O <-> H2 + 1/2 O2}, to introduce \ce{H2} into the reaction. The potential-dependent energy of formation for each oxidized particle, $\Delta E_\text{F}^\text{system}(\Delta \phi)$, can then be calculated as
\begin{eqnarray}
    \Delta E_\text{F}^\text{system}(\Delta \phi) = E_\text{tot}^\text{system} - E_\text{tot}^\text{ref} - N_\text{O}\ \left(\mu_\text{O} + 2 e\, \Delta \phi \right)\ \text{with} \\
    \mu_\text{O} = \mu_\text{\ce{H2O}} - \mu_\text{\ce{H2}} ,
\end{eqnarray}
where $E_\text{tot}^\text{system}$ is the computed total energy of the oxidized system, $E_\text{tot}^\text{ref}$ is the total energy of a reference system (here: the pristine NP without adatoms), $N_\text{O}$ is the number of oxygen adatoms in the oxidized system, $\mu_\text{O}$, $\mu_\text{\ce{H2O}}$, and $\mu_\text{\ce{H2}}$ are the chemical potentials of oxygen, water, and hydrogen, respectively, $e$ is the elementary charge, and $\Delta \phi$ is the electrode potential. The resulting electrochemical phase diagram, which shows the thermodynamically most stable oxide structures at SHE conditions, is given in Fig. \ref{fig:echemphasediagram}.
The coverage convention introduced in section \ref{s:coverage} is used to classify the oxidized structures.

A code example for constructing electrochemical phase diagram from a dataset of oxidized Pt NPs can be found in the supporting git repository in \verb|phaseDiagram.py|.

\section{Conclusions}
The present work reviewed analytical methods useful for extracting thermodynamic information from computational simulations of NP systems. As computational methods continue to improve and costs for high-performance computing are brought down by the year, simulations of large NP model systems instead of stand-ins such as periodic surface models will become the norm. Many established analytical methods, such as partial charge analysis or radial distribution functions, can easily be scaled for application to NP systems. On the other hand, the more complex nature of NPs requires new methods and approaches to analyze structure-activity relationships in particular. The authors hope that the present work will serve as a starting point for computational researchers working with NP structures.

\section*{Acknowledgement}

BK acknowledges funding from the Professional Graduate and Training Center Ulm (ProTrainU). Further, the authors gratefully acknowledge support by the DFG (german science foundation) through the collaborative research centers SPP-2248 (project ID 441209207) as well as SFB-1279 (project ID 316249678).

CRediT author contribution statement. BK: conceptualization, methodology, data curation, visualization, writing - original draft. CJ: writing - review \& editing. DG: writing - review \& editing. DF: conceptualization, supervision. HJ: supervision. TJ: writing - review \& editing, supervision, funding acquisition.

\section*{Supporting Information}

An online, Jupyter book based tutorial is available which outlines how to implement many of the methods presented here: \href{https://bjk24.gitlab.io/in-silico-review/intro.html}{https://bjk24.gitlab.io/in-silico-review/intro.html}. The code is built to interact with a recently published data set of oxidized NP structures (DOI: \href{https://doi.org/10.5281/zenodo.6322004}{10.5281/zenodo.6322004}).\cite{Kirchhoff2022} All figures in this manuscript involving the oxidized 3 nm octahedral NPs were obtained using these scripts and data set. We recommend interested readers to execute and modify the provided code examples to deepen their understanding of these various evaluation methods.




\bibliography{insilico-characterization.bib} 
\bibliographystyle{rsc} 

\end{document}